\documentclass[aps,prb,floatfix,epsfig,twocolumn,showpacs,preprintnumbers]{revtex4}
\usepackage{amsmath,amssymb,graphicx,bm,epsfig}
\usepackage{color}
\usepackage{braket}
\usepackage{hyperref}

\newcommand{\vR}{{\mathbf{R}}}
\newcommand{\vF}{{\mathbf{F}}}
\renewcommand{\vr}{{\mathbf{r}}}

\newcommand{\vk}{{\mathbf{k}}}

\newcommand{\vK}{{\mathbf{K}}}
\newcommand{\vG}{{\mathbf{G}}}
\newcommand{\vq}{{\mathbf{q}}}

\newcommand{\vS}{{\mathbf{S}}}

\newcommand{\cO}{{\cal O}}
\newcommand{\cH}{{\cal H}}
\newcommand{\cV}{{\cal V}}

\newcommand{\cU}{{\cal U}}

\newcommand{\Tr}{\mathrm{Tr}}

\renewcommand{\Im}{\textrm{Im}}

\newcommand{\vcA}{\vec{\cal A}}
\newcommand{\vcB}{\vec{\cal B}}
\newcommand{\vcC}{\vec{\cal C}}
\newcommand{\ovcA}{\vec{\overline{\cal A}}}
\newcommand{\ovcB}{\vec{\overline{\cal B}}}
\newcommand{\ovcC}{\vec{\overline{\cal C}}}
\newcommand{\cB}{{\cal B}}

\begin{document}

\title{Forces for Structural Optimizations in Correlated Materials within
  DFT+Embedded DMFT Functional Approach}

\author{Kristjan Haule and Gheorghe L. Pascut}
\affiliation{Department of Physics, Rutgers University, Piscataway, NJ 08854, USA}
\date{\today}

\begin{abstract}
  We implemented the derivative of the free energy functional with
  respect to the atom displacements, so called force, within the
  combination of Density Functional Theory and the Embedded Dynamical
  Mean Field Theory. We show that in combination with the numerically
  exact quantum Monte Carlo (MC) impurity solver, the MC noise cancels
  to a great extend, so that the method can be used very efficiently
  for structural optimization of correlated electron materials. As an
  application of the method, we show how strengthening of the
  fluctuating moment in FeSe superconductor leads to a substantial
  increase of the anion height, and consequently to a very large
  effective mass, and also strong orbital differentiation.
\end{abstract}
\pacs{71.27.+a,71.30.+h}
\date{\today}
\maketitle

\section{Introduction}

The theoretical crystal structure prediction is one of the most
fundamental challenge  in condensed matter physics and material science,
but it was not until 90s that computers became sufficiently powerful
to allow predictions of crystal structures from first principles of
very simple materials.~\cite{Tsuneyuki,Maddox} The last decade has
witnessed a tremendous advance in our ability to predict crystal
structures from ab-initio, mostly due to the development of efficient
minimization algorithms for finding minimums in complex total energy
landscape of solids~\cite{PreUspex,Uspex,likeUspex}, and because of prior
development of efficient implementations of the Density Functional
Theory (DFT) methods. The core of almost all these algorithms is based on the DFT
stationary functional, which delivers the total energy of the solid
and the forces on all atoms in the unit cell. 
However DFT, in its semilocal approximations such as the local density
approximation (LDA) or generalized gradient approximation (GGA),
fails to predict the ground state of many correlated electron
materials, such as the Mott insulators and correlated metals,
therefore the crystal structure predictions in such systems are
severely hampered by inaccuracy of available DFT functionals.

It is well known that the DFT total energies are many times
surprisingly good, even when the electronic structure is completely
wrong, such as for example in high-Tc cuprates. This is because the
DFT total energy functional is stationary, i.e., the first derivative
of the energy with respect to electronic charge vanishes. Therefore a
relatively small reorganization of the low energy valence charge
density gives not too large correction to the total energy.

There are nevertheless many documented failures of LDA and GGA in
predicting crystal structures of correlated materials such as in Ce
metal, Pu, and transition metal oxides such as FeO. For the Hund's
metals~\cite{Haule_njp,Yin-nm11}, such as the iron superconductors,
the pnictogen height is grosly underestimated by DFT for
about 0.15$\textrm{\AA}$.

To account for the correlation effects beyond semi-local
approximations of DFT, more sophisticated many body methods have been
developed. Among them, one of the most successful algorithms is the
combination of the dynamical mean-field theory (DMFT) and
DFT~\cite{KotliarFirst,Lichtenstein,rmp2}, which is also based on the idea
of locality of correlations, but in the case of DMFT only the locality
of correlations to a given atom is explored, which is much less
restrictive than locality to a point in 3D space in DFT semi-local
approximations. This DFT+DMFT method has achieved great success in
numerous correlated materials (for a review see Ref.~\onlinecite{rmp2}), but its
potential for structural optimization has not been much explored. This
is mostly because the majority of the implementations of this method
are not implementing the DFT+DMFT functional. Instead they typically
build the low energy model first, and then solve the Hubbard-like
model by the DMFT method, thus losing the stationarity property, and
hence the precision of the resulting total energies.

The stationary implementation of the DFT+embedded DMFT functional has
been achieved recently~\cite{FreeE}, which opened the possibility of
computing forces to high-enough precision for theoretical optimization
of structures. The present manuscript details how this is achieved
very efficiently within all electron Linearized Augmented Plane-wave
(LAPW) implementation.

We will also show that in combination with the Quantum Monte Carlo
(QMC) impurity solver, the forces can be converged to even higher
accuracy than the free energy itself, which seems surprising at first,
as only the free energy is stationary, while the forces are not. But
as explained below, this is because some quantities can be more
accurately computed by QMC than others.  As QMC method has
inherent statistical noise, such noise cancelation in computing forces
is very wellcome and extremely useful for practical implementations.

The reason that the free energy is hard to compute by the exact QMC
impurity solver, is that it is not possible to accurately sample the
interacting part of the free energy functional, the so-called
Baym-Kadanoff functional $\Phi[G]$. Essentially, $\Phi[G]$ contains
the entropy of the system, which is notoriously hard to compute within
the Monte Carlo methods.~\cite{Simulations}  An alternative approach was invented in
Ref.~\onlinecite{FreeE}, which still requires integration over
temperature for the entropy term. However, as we will show below, the
force requires only the first derivative $\delta\Phi[G]/\delta G$,
which is the familiar self-energy $\Sigma$, and which can be computed
to very high accuracy in QMC method. It turns out that only the first
derivative of the free energy functional, i.e., the force, can be so
accurately implemented. To compute the free energy itself, one needs
$\Phi[G]$, which is hard to compute. For the phonon spectra, which is
the second derivative, one needs $\delta^2 \Phi[G]/\delta G^2$, which
is the two particle vertex, and is again very hard to accurately
compute in practice. Therefore only the force on atoms can be computed very
precisely in the DFT+embedded DMFT functional (DFT+EDMFTF) method when
the exact QMC method is used as the impurity solver.

As a consequence, the frozen phonon approach is more tractable than
the generalization of the density functional perturbation
theory~\cite{DFTPerturb}. Also the integration of the force will
likely be the best way to calculate phase diagrams of correlated
solids, as the force can be converged to much higher precision than
the free energy itself.

We are aware of two prior reports on computing forces and other
derivatives within DFT+DMFT method. The work of Savrasov and
Kotliar~\cite{SavrasovLR} considered only the second derivative of the
DFT+DMFT functional with respect to atom displacement, to obtain the
phonon spectra. They considered only the finite wave vector $\vq$, to
avoid the need of differentiating the Kohn-Sham eigen-energies, which
are needed for evaluating the forces.  Moreover, using the Hubbard-I
impurity solver, they also neglected the change of the DMFT
self-energy with respect to the atom displacement
($\delta\Sigma/\delta G=\delta^2\Phi/\delta G^2$), which plays an
important role in our method.  The work of Leonov~\textit{et. al.}
~\cite{Anisimov_Force} reported computation of forces within DFT+DMFT,
however, their implementation is not based on stationary
functional. The derivative of non-stationary DMFT total energy was
computed, in which the two-particle vertex is needed at all
frequencies, which is extremely hard to compute accurately enough by
the present day impurity solvers, to be useful for the structural
optimizations.  Moreover, the method of
Leonov~\textit{et. al.}~\cite{Anisimov_Force} is a based on the two
step process, where the low energy model is build first and then a
Hubbard model is solved by the DMFT method. Also the influence of the
DMFT correlations on the electronic charge, needed in the DFT step, is
usually neglected. These two approximations are a source of inaccuracy, which
is hard to overcome, even when the impurity is solved with a very high
precision so that the two-particle vertex is converged within $meV$
accuracy. Hence alternative approaches are needed for practical
predictions of crystal structures for correlated electron solids.

The manuscript is organized as follows: In Section~\ref{Chap1} we
derive the equations for the forces within DFT+Embedded DMFT
functional. In part~\ref{Chap1a} we introduce the Luttinger-Ward
functional and its derivative with respect to the atom displacement,
which is the well known Hellmann-Feynman force. In part~\ref{Chap1b}
we derive a basis set independent expression for the Pulay force, the
additional force due to basis set discretization. In part~\ref{Chap1c}
we show how is this formula evaluated in a mixed basis set, in which
the basis has both the atom-centered and origin-less functions. In
part~\ref{Chap1d} we derive Pulay forces in one such basis, namely the
LAPW basis. In chapter~\ref{Chap2} we apply this method to FeSe, and
show how quantum Monte Carlo noise cancels to large extent when
computing the force. 
In chapter~\ref{Chap2} we also show that FeSe is positioned in
the critical region where a small increase of the fluctuating moment
on Fe leads to substantial increase of Se-height, and consequently
also of the correlation strength.  In appendix~\ref{AppendixA} we give
details of the force evaluation within the LAPW basis set.

\section{Derivation of the Force within DFT+EDMFTF}
\label{Chap1}

The force on an atom is defined as minus the change of the total free
energy when its nucleus is displaced by a small amount.  The
Hellmann-Feynman theorem~\cite{HFT} states that this force is equal to
the electrostatic force on the nucleus, but due to discretization of
the problem, which involves convenient atom centered basis
and atom centered projector, the actual force on an atom has
additional contributions, which are usually called Pulay
forces~\cite{Pulay}.

\subsection{The Luttinger-Ward approach}
 \label{Chap1a}

In ab-initio electronic structure methods, the force is computed by
evaluating the analytical derivative of the total energy functional.
In order to compute such derivatives, it is very convenient to use a
stationary functional, in which a small change of the electron density
(and the Green's function), leaves functional invariant.  Indeed, if
the implementation of the functional is exact, one could evaluate the
force by considering a small displacement of nuclei at fixed
electron charge density (and fixed Green's function). Namely, the
total derivative of the free energy functional $\Gamma[G]$ can be
split into two terms, the partial derivatives with respect to the
Green's function at fixed atomic positions, and the partial
derivatives with respect to displacements at fixed Green's function,
i.e.,
\begin{eqnarray}
\frac{\delta\Gamma[G]}{\delta\vR_\mu} = 
\left(\frac{\partial\Gamma[G]}{\partial \vR_\mu}\right)_{G}
+\int d\vr d\vr'\frac{\delta G(\vr\vr')}{\delta\vR_\mu}\left(\frac{\partial\Gamma[G]}{\partial G(\vr\vr')}\right)_{R_\mu}
\end{eqnarray}
If the functional is stationary, it follows that
$\left(\frac{\partial\Gamma[G]}{\partial G}\right)_{R_\mu}=0$, and
therefore only the first term contributes, and gives so-called Hellmann-Feynman forces.

In the Green's function approaches, such as the Dynamical Mean Field
Theory, the free energy functional is best expressed by the stationary
Luttinger-Ward functional, which takes the form
\begin{eqnarray}
\Gamma[G] = \Tr\log(-G) - \Tr((G_0^{-1}-G^{-1})G) 
\nonumber\\
+ \Phi[G]+   E_{nuclei}
\label{Eq:Gamma}
\end{eqnarray}
Here $\Tr$ runs over spatial degrees of freedom, the spin, and when
quantities are dynamic, also over Matsubara frequencies.
Note that the derivative with respect to the Green's function at constant
ion position $\left(\frac{\partial\Gamma[G]}{\partial  G}\right)_{R_\mu}$ is 
$G^{-1}-G_0^{-1} + \frac{\delta\Phi[G]}{\delta G}$,
and as expected vanishes, because the system satisfies the Dyson
equation $G^{-1}=G_0^{-1}-\frac{\delta\Phi[G]}{\delta G}$.  The
only term that explicitly depends on the nucleus position is contained
in $G^0$ and $E_{nuclei}$, and the force thus becomes
\begin{eqnarray}
\frac{\delta\Gamma[G]}{\delta\vR_\mu} =-\Tr(G\frac{\partial
  G_0^{-1}}{\partial\vR_\mu})+\frac{\partial E_{nuclei}}{\vR_\mu}
\nonumber\\
=\Tr(\rho \frac{\partial V_{nuclei}}{\partial\vR_\mu})+\frac{\partial E_{nuclei}}{\vR_\mu}
\label{Eq:dGamma}
\end{eqnarray}
where $G_0^{-1}=i\omega_n+\mu-T-V_{nuclei}$, and $T$, $V_{nuclei}$ are the kinetic
energy operator and the potential due to nuclei, respectively. Because
$V_{nuclei}$ is frequency independent, we performed a partial trace
over Matsubara frequency to replace the Green's function with the charge
density in the first term $\Tr(G\delta V_{nuclei})=\Tr(\delta
V_{nuclei}\frac{1}{\beta}\sum_{i\omega_n}G(i\omega_n))=\Tr(\rho V_{nuclei})$.
The derivative in Eq.~\ref{Eq:dGamma} then gives 
\begin{eqnarray}
\vF^{HF}=-\Tr(\rho \frac{\partial V_{nuclei}}{\partial\vR_\mu})-\frac{\partial E_{nuclei}}{\vR_\mu},
\label{Eq:HFF}
\end{eqnarray}
which is the Hellmann-Feynman force.

\subsection{Forces within DFT+EDMFTF approach}
\label{Chap1b}
The exact Baym-Kadanoff $\Phi$ functional is the sum of all skeleton
Feynman diagrams, which can not be computed exactly for the solid
state systems we are interested in. Within DFT+embedded
DMFT functional (DFT+EDMFTF) approach, the $\Phi$ functional is approximated by
the following superposition of terms
\begin{eqnarray}
\Phi[G] = E_H[\rho] + E_{xc}[\rho] + \sum_{\vR_\mu}\Phi^{DMFT}[G^\mu_{loc}] -  \Phi^{DC}[\rho^\mu_{loc}] 
\label{Eq:PhiDMFT}
\end{eqnarray}
Here the first two terms give rise to usual DFT equations, the third
term adds all Feynman diagrams, local to selected set of atoms at
$\vR_\mu$.  The last term subtracts the interaction, which is
accounted for by both approximations. The latter is now known
exactly.~\cite{exactDC}

Notice that $\Phi^{DMFT}[G^\mu_{loc}]$ has the same functional form as
the exact functional $\Phi^{exact}_{V_C}[G]$, however, to obtain
$\Phi^{DMFT}$ from $\Phi^{exact}_{V_C}[G]$, the Green's function $G$
is truncated to its local component $G\rightarrow G_{loc}$, and
Coulomb correlation $V_{C}$ is screened, due to this truncation. Such
truncation of variable of interest parallels the LDA and GGA type
approximation to DFT, where $E_{XC}$ is similarly taken to be local
(semilocal) to each point in 3D space, which is clearly a more
restrictive approximation. The combined DFT+EDMFTF is thus a good
compromise between speed and accuracy, as most of the degrees of freedom
are treated on semilocal level, while the correlated orbitals are
augmented by the best local approximation to a given correlated atom.
Notice also that it is possible to define somewhat different
functional $\Gamma$, which gives the exact local Green's function and
the exact free energy in its stationary point~\cite{Chitra}, and for
which the diagrammatic rules were also developed in Ref.~\onlinecite{Chitra}.  In
practice, however, a successful approximation that would go beyond DMFT and
would not add an exponential cost (like cluster extensions) has not been
developed yet from this formalism.

To define the ``locality to an atom'' in Eq.~\ref{Eq:PhiDMFT}, we need
to define the DMFT projector, and in the embedded DMFT approach, this
projector is chosen to be a set of atom centered functions
$\ket{\phi^\mu_m}$, so that
\begin{eqnarray}
G^\mu_{loc}(\vr,\vr') = \sum_{mm'}\braket{\vr|\phi^\mu_m}\braket{\phi^\mu_m|G|\phi^\mu_{m'}}\braket{\phi^\mu_{m'}|\vr'}.
\end{eqnarray}
If these functions $\ket{\phi^{\mu}_m}$ form a complete basis, then DMFT
method is projector independent, except that it dependents on the range
of the projector (the sphere size). In practice, the solutions of
the radial Schroedinger equation that correspond to the $3d$, $2p$,
and $4s$ solutions, of say an Fe atom, are sufficiently separated in
energy so that only $3d$ states need to be treated dynamically, while
the rest of the orbitals can safely be treated statically within
the exchange-correlation approximation.

The stationarity of the functional $\Gamma[G]$, when using $\Phi[G]$
of the DFT+EDMFTF (Eq.~\ref{Eq:PhiDMFT}), gives the Dyson equation
\begin{eqnarray}
G^{-1}-G_0^{-1} + (V_H+V_{xc})\delta(\vr-\vr')\delta(\tau-\tau')
\nonumber\\
 +  \sum_{m m',\vR_\mu}\braket{\vr|\phi^\mu_m}\braket{\phi^\mu_m|\Sigma-V_{DC}|\phi^\mu_{m'}}\braket{\phi^\mu_{m'}|\vr'} =0,
\label{Eq:Dysn}
\end{eqnarray}
hence the electron Green's function must satisfy
\begin{eqnarray}
G^{-1}=i\omega_n+\mu-T-(V_{nuclei} +V_H+V_{xc})-
\nonumber\\
  \sum_{mm'\vR_\mu}\ket{\phi^\mu_m}\braket{\phi^\mu_m|\Sigma_{i\omega_n}-V_{DC}|\phi^\mu_{m'}}\bra{\phi^\mu_{m'}}
\end{eqnarray}
and the functional $\Gamma[G]$ reaches extremum for this $G$.  When
inserting extremal $G$ back into $\Gamma[G]$ (Eq.~\ref{Eq:Gamma}), the
value of $\Gamma$ gives the free energy of the system~\cite{AGDbook},
which hence becomes
\begin{eqnarray}
F = \Tr\log\left(-G\right) 
- \Tr((V_H+V_{xc})\rho) 
+ E_{H}[\rho]
\nonumber\\
+E_{xc}[\rho] 
+ E_{nuclei}
- \Tr((\Sigma-V_{DC})\braket{\phi|G|\phi}) 
\nonumber\\
+ \sum_{\vR_\mu}\Phi^{DMFT}[G^\mu_{loc}]-\Phi^{DC}[\rho^\mu_{loc}] 
+\mu N
\label{DFMT:func0}
\end{eqnarray}
Notice that $(\braket{\phi|G|\phi})_{mm'}$ are the matrix elements of the
local Green's function $\braket{\phi_m^\mu|G|\phi_{m'}^\mu}$.

In the all-electron calculations of the free energy, the spatial
degrees of freedom are expanded in terms of a mixed basis set, which
includes atom centered basis functions, therefore the Hellmann-Feynman
force is very different from the derivative of the implemented free
energy Eq.~\ref{DFMT:func0}. It is therefore essential to find the analytic derivative of
the actually implemented free energy Eq.~\ref{DFMT:func0}. This is
derived below. We will concentrate on the valence electron
contribution, as the core contribution within DFT+EDMFTF is the same
as in DFT.

To evaluate the logarithm of the Green's function in Eq.~\ref{DFMT:func0}, we first solve the following
frequency dependent eigenvalue-problem
\begin{widetext}
\begin{eqnarray}
\bra{\psi_{j\vk\omega_n}}(T+ V_{nuclei} +V_H+V_{xc}+
\sum_{mm'\vR_\mu}\ket{\phi^\mu_m}\braket{\phi^\mu_m|\Sigma_{i\omega_n}-V_{DC}|\phi^\mu_{m'}}\bra{\phi^\mu_{m'}}
)\ket{\psi_{i\vk\omega_n}}=\delta_{ij}\;\varepsilon_{\vk\omega_n,i}
\label{Eq:eigenval}
\end{eqnarray}
\end{widetext}
so that the Green's function is simply given by
\begin{eqnarray}
\braket{\psi_{j\vk\omega_n}|G|\psi_{i\vk\omega_n}} = \frac{\delta_{ij}}{i\omega_n+\mu-\varepsilon_{\vk\omega_n,i}}
\label{Eq:gij}
\end{eqnarray}
and the free energy is evaluated by
\begin{eqnarray}
F = -\Tr\log\left(- i\omega_n-\mu+\varepsilon_{\vk\omega_n}\right) 
- \Tr((V_H+V_{xc})\rho) 
\nonumber\\
+ E_{H}[\rho]+E_{xc}[\rho] 
+ E_{nuclei}
- \Tr((\Sigma-V_{DC})\braket{\phi|G|\phi}) 
\nonumber\\
+ \sum_{\vR_{\mu}}\Phi^{DMFT}[G^\mu_{loc}]-\Phi^{DC}[\rho^\mu_{loc}] 
+\mu N
\label{DFMT:func}
\end{eqnarray}
This is the actual expression implemented in DFT+EDMFTF code. To get
the force on an atom, we need to consider a small variation of this
energy when moving an atom at position $\vR_\mu$
\begin{eqnarray}
\delta F = \Tr\left(\frac{\delta\varepsilon_{\vk\omega_n}-\delta\mu}{i\omega_n+\mu-\varepsilon_{\vk\omega_n}}\right)
-\Tr(\rho(\delta V_H+\delta V_{xc}))
\nonumber\\
-\Tr(G_{loc} (\delta\Sigma-\delta V_{DC})) + \delta E_{nuclei}+N\delta\mu 
\end{eqnarray}
where we used the fact that
\begin{eqnarray}
&&\delta(E_{H}+E_{xc}) = \Tr((V_H+V_{xc})\delta\rho)\\
&&\sum_{\vR_\mu}\delta \Phi^{DMFT}[G^\mu_{loc}]+\delta\Phi^{DC}[\rho^\mu_{loc}] = \Tr((\Sigma-V_{DC}) \delta G_{loc})
\nonumber
\end{eqnarray}
and, as we work at constant electron density, $\delta N=0$. Inserting
the Hellmann-Feynman forces Eq.~\ref{Eq:HFF}, we arrive at
\begin{eqnarray}
\delta F = \Tr\left(\frac{\delta\varepsilon_{\vk\omega_n}}{i\omega_n+\mu-\varepsilon_{\vk\omega_n}}\right)
-\Tr(\rho\;\delta V_{KS})
\nonumber\\
-\Tr(G_{loc} (\delta\Sigma-\delta V_{DC}))-\sum_\mu \vF^{HF}_\mu\delta \vR_\mu
\label{Eq:deltaF}
\end{eqnarray}
where $V_{KS} = V_H+V_{xc}+V_{nuclei}$.

Finally, we define the Pulay force on an atom $\vF^{Puly}$ as the addition
to the Hellmann-Feynman force (due to the basis set in which the
functional is implemented)
$\delta F=-\sum_\mu (\vF^{HF}_\mu+\vF^{Puly}_\mu)\delta \vR_\mu$.
From Eq.~\ref{Eq:deltaF} it follows that
the Pulay forces are
\begin{eqnarray}
\vF^{Puly}_\mu &=&
-\Tr\left(\frac{1}{i\omega_n+\mu-\varepsilon_{\vk\omega_n}}\frac{\delta\varepsilon_{\vk\omega_n}}{d\vR_\mu}\right) 
\nonumber\\
&+&\Tr\left(\rho\frac{\delta V_{KS}}{\delta\vR_\mu}\right)
+\Tr\left(G_{loc} \frac{\delta\Sigma-\delta V_{DC}}{\delta\vR_\mu}\right)
\label{Eq:Fpulley}
\end{eqnarray}
This equation is still completely general expression for the force within the
DFT+EDMFTF, irrespectively of the basis set employed.


\subsection{Pulay forces expressed in a mixed basis set}
\label{Chap1c}

To proceed, we need to choose a basis to express the electron Green's
function. We will here denote it by $\ket{\chi_\vK}$, (as we have in
mind LAPW basis set) but the details of the basis are not important
here, so this derivation is relevant for any mixed basis set.

The DMFT eigenvectors $\ket{\psi_{i\vk\omega_n}}$
are than expanded in the chosen basis in the usual way
\begin{eqnarray}
\ket{\psi_{i\vk\omega_n}} = \sum_{\vK} \ket{\chi_{\vK}}A^R_{\vK i}
\label{Eq:exp1}
\\
\bra{\psi_{i\vk\omega_n}} = \sum_{\vK} A^L_{i\vK}\bra{\chi_{\vK}}
\label{Eq:exp2}
\end{eqnarray}
Note that the eigenvectors $\ket{\psi_{i\vk\omega_n}}$ are momentum
and frequency dependent, hence $A^R_{\vK i}$ also inherit this momentum
and frequency dependence, i.e., $A^R_{\vK i}=A^R_{\vK  i}(\vk,\omega_n)$. Note also that the eigenvalue problem is not
Hermitian, therefore we need to distinguish between the right and the left eigenvectors.
Using expansion Eqs.~\ref{Eq:exp1} and~\ref{Eq:exp2}, the DMFT eigenvalue problem Eq.~\ref{Eq:eigenval} reads
\begin{eqnarray}
\sum_{\vK\vK'}A^L_{j\vK'} \left[H^0_{\vK'\vK} + V_{\vK'\vK}\right]A^R_{\vK i}=\delta_{ij}\;\varepsilon_{\vk\omega_n,i}
\label{Eq:eigenvalue}
\end{eqnarray}
where
\begin{eqnarray}
&& H^0_{\vK'\vK}=\braket{\chi_{\vK'}|T+ V_{nuclei}   +V_H+V_{xc}|\chi_\vK}
\label{def:V}\\
&& V_{\vK'\vK}= \sum_{mm'\vR_\mu}\braket{\chi_{\vK'}|\phi^\mu_m}\braket{\phi^\mu_m|\Sigma-V_{DC}|\phi^\mu_{m'}}\braket{\phi^\mu_{m'}|\chi_\vK}
\nonumber
\end{eqnarray}
Here $H^0$ stands for the DFT part of the Hamiltonian, and $V$ for the
additional DMFT contributions.

The eigenvectors are orthogonalized in the usual way 
$$\sum_{\vK\vK'}A^L_{i\vK'} O_{\vK'\vK} A^R_{\vK j} = \delta_{ij}$$
where $O_{\vK'\vK}=\braket{\chi_{\vK'}|\chi_{\vK}}$ is the overlap
matrix, hence the eigenvalue problem Eq.~\ref{Eq:eigenvalue} can be cast in the following form
\begin{eqnarray}
\sum_{\vK}\left[H^0_{\vK'\vK} + V_{\vK'\vK}\right]A^R_{\vK i}= \sum_{\vK}O_{\vK'\vK} A^R_{\vK i}\;\varepsilon_{\vk\omega_n,i}
\label{Eq:eigenvalue2}
\end{eqnarray}
or in short notation
$$[H^0 + V]A^R = O A^R \varepsilon.$$
Eq.~\ref{Eq:eigenvalue2} is enforced for any position of atoms $\vR_\mu$, hence its variation vanishes.
We thus have
\begin{eqnarray}
[(\delta H^0) + (\delta V)]A^R + [H^0+V]\delta A^R 
\nonumber\\
= (\delta O) A^R \varepsilon + O (\delta A^R) \varepsilon + O A^R \delta\varepsilon
\end{eqnarray}
and multiplying with $A^L$ we get
\begin{eqnarray}
A^L [(\delta H^0) + (\delta V)]A^R + A^L [H^0+V]\delta A^R 
\nonumber\\
=A^L (\delta O) A^R \varepsilon + A^L O (\delta A^R) \varepsilon +\delta\varepsilon
\end{eqnarray}
We also use the fact that $A^L[H^0+V]=\varepsilon A^L O$ to obtain
\begin{eqnarray}
\delta\varepsilon = A^L [(\delta H^0) + (\delta V)]A^R -A^L (\delta O)  A^R \varepsilon 
\nonumber\\
+ \varepsilon A^L O(\delta A^R) - A^L O (\delta A^R) \varepsilon 
\end{eqnarray}
In Eq.~\ref{Eq:Fpulley} we only need the diagonal variation of the eigenvalues $(\delta\varepsilon)_{ii}$, for which the last two terms cancel because $\varepsilon$ is diagonal matrix, hence $\varepsilon_i (A^L O(\delta A^R))_{ii} - (A^L O (\delta A^R)_{ii} \varepsilon_i=0$. We thus obtain
\begin{eqnarray}
(\delta\varepsilon_{\vk\omega_n})_{ii} = \sum_{\vK\vK'} A^L_{i\vK'}  [\delta H^0_{\vK'\vK} + \delta V_{\vK'\vK}]A^R_{\vK i} 
\nonumber\\
-A^L_{i\vK'}\, \delta O_{\vK'\vK} \, A^R_{\vK i}\, \varepsilon_{\vk\omega_n,i}
\end{eqnarray}
This is a dynamic generalization of the DFT expression, derived in
Ref.~\onlinecite{Krakauer}.

Next we split the DMFT eigenvectors into the static (Kohn-Sham) part, and the frequency dependent part
\begin{eqnarray}
A^R_{\vK i} = \sum_j A^0_{\vK j} (B^R_{\omega_n})_{ji}\\
A^L_{i\vK} = \sum_j (B^L_{\omega_n})_{ij}A^{0\,\dagger}_{j\vK}
\end{eqnarray}
or short
$A^R = A^0 B^R_{\omega_n}$ and $A^L=B^L_{\omega_n} A^{0\dagger}$.
Here $A^0$ satisfies the Kohn-Sham eigenvalue problem $A^{0\dagger} H^0 A^0=\varepsilon^0$.

In terms of the above defined quantities Eq.~\ref{Eq:Fpulley} takes
the form
\begin{widetext}
\begin{eqnarray}
\vF^{Puly}_\mu = 
-\Tr\left(G^d B_{\omega_n}^L 
\left[
A^{0\dagger}\left(\frac{\delta H^0}{\delta\vR_\mu}+\frac{\delta V}{\delta\vR_\mu}\right)A^0 B_{\omega_n}^R
-A^{0\dagger}\frac{\delta O}{\delta\vR_\mu} A^0 B_{\omega_n}^R\varepsilon_{\vk\omega_n}
\right]
\right)
+\Tr\left(\rho\frac{\delta V_{KS}}{\delta\vR_{\mu}}\right)+\Tr\left(G_{loc} \frac{\delta\Sigma-\delta V_{DC}}{\delta\vR_\mu}\right)
\label{Eq:deltaF2}
\end{eqnarray}
\end{widetext}
where we denoted
$$G^d=\frac{1}{i\omega_n+\mu-\varepsilon_{\vk\omega_n}},$$
and $G^d$ is the Green's function in diagonal representation.
Next we define the following DMFT density matrices
\begin{eqnarray}
\widetilde{\rho}  &\equiv& \frac{1}{\beta}\sum_{i\omega_n} B^R_{\omega_n}\frac{1}{i\omega_n+\mu-\varepsilon_{\vk\omega_n}}B^L_{\omega_n}\\
\widetilde{(\rho\varepsilon)} &\equiv& \frac{1}{\beta}\sum_{i\omega_n} B^R_{\omega_n}\frac{\varepsilon_{\vk\omega_n}}{i\omega_n+\mu-\varepsilon_{\vk\omega_n}}B^L_{\omega_n}
\end{eqnarray}
which are the usual DMFT density matrices, but here written in the
Kohn-Sham basis. 
Note that the density matrix
$\widetilde{\rho}$ can also be expressed by
$\widetilde{\rho}_{ij}=\braket{\psi^0_i|\rho|\psi^0_j}$
where $\ket{\psi^0}$ are Kohn-Sham eigenvectors of $H^0$ and $\rho$ is
the self-consistent charge density of DFT+EDMFTF method.
We also recognize the Green's functions written in the $\ket{\chi_\vK}$ basis
\begin{eqnarray}
\bar{G}_{\vK\vK'} = (A^0 B_{\omega_n}^R\frac{1}{i\omega_n+\mu-\varepsilon_{\vk\omega_n}} B_{\omega_n}^L A^{0\dagger})_{\vK\vK'}
\label{Eq:GLAPW}
\end{eqnarray}
The overline here is used to stress that the Green's function is
expressed in the basis of $\ket{\chi_\vK}$ (rather than in real space).
This allows us to simplify 
\begin{widetext}
\begin{eqnarray}
\vF^{Puly}_\mu = -\Tr\left(\widetilde{\rho} A^{0\dagger}\frac{\delta  H^0}{d\vR_\mu}A^0 -\widetilde{(\rho\varepsilon)} A^{0\dagger}\frac{\delta  O}{\delta\vR_\mu} A^0 \right) +
\Tr\left(\rho\frac{\delta V_{KS}}{\delta\vR_\mu}\right)
-\Tr\left(\bar{G}\frac{\delta V}{\delta\vR_\mu} \right) 
+\Tr\left(G_{loc} \frac{\delta\Sigma-\delta V_{DC}}{\delta\vR_\mu}\right)
\label{Eq:deltaF3}
\end{eqnarray}
\end{widetext}

We next simplify the interacting part (the third term above), which
contains interaction $V$ (defined by Eq.~\ref{def:V}):
\begin{eqnarray}
&&\Tr\left(\bar{G}\delta V \right)
\nonumber\\
&&  =\frac{1}{\beta}
\sum_{\substack{i\omega,m'm\\\vK\vK'}}
\bar{G}_{\vK\vK'}\delta\left(\braket{\chi_{\vK'}|\phi_{m'}}(\Sigma-V_{DC})_{m'm}\braket{\phi_m|\chi_{\vK}}\right)
\nonumber\\
&&=
\frac{1}{\beta}\sum_{\substack{i\omega_n,m'm\\\vK\vK'}} \bar{G}_{\vK\vK'}
(\Sigma-V_{DC})_{m'm}\delta\left(\braket{\chi_{\vK'}|\phi_{m'}}\braket{\phi_m|\chi_{\vK}}\right)
\nonumber\\
&&+\Tr\left(G_{loc}(\delta\Sigma-\delta V_{DC})\right)
\end{eqnarray}
where we used the fact that
$$(G_{loc})_{mm'}=\sum_{\vK\vK'}\braket{\phi_m|\chi_{\vK}}\bar{G}_{\vK\vK'}\braket{\chi_{\vK'}|\phi_{m'}}$$
Finally, the Pulay forces become
\begin{widetext}
\begin{eqnarray}
\vF^{Puly}_\mu =-\Tr\left(\widetilde{\rho} A^{0\dagger} \frac{\delta H^0}{\delta\vR_\mu}A^0 -\widetilde{(\rho\varepsilon)} A^{0\dagger}\frac{\delta O}{\delta\vR_\mu} A^0 \right) +\Tr\left(\rho\;\frac{\delta V_{KS}}{\delta\vR_\mu}\right)
\nonumber\\
-\frac{1}{\beta}\sum_{i\omega_n}\sum_{\vK\vK',m'm} \bar{G}_{\vK\vK'}(\Sigma-V_{DC})_{m'm}\frac{\delta\left(\braket{\chi_{\vK'}|\phi_{m'}}\braket{\phi_m|\chi_{\vK}}\right)}{\delta\vR_\mu}
\label{Eq:Pulay}
\end{eqnarray}
\end{widetext}
This is still a basis independent expression of the Pulay force,
as we abstain discussing specifics of a given basis set, but  we
nevertheless managed to avoid the expensive frequency summations in all but the
last term. To perform the expensive $\vK$ and frequency summation in
the last term, we need to determine the derivative of the projector,
which depends on the basis set and the choice of a projector.

\subsection{Pulay forces within LAPW basis and quasi atomic orbital projector}
\label{Chap1d}

Within the LAPW method~\cite{APW,LAPW} the interstitial space is
spanned by the plane waves $\widetilde{\chi}_\vK$, while inside the
muffin-tin spheres, the plane waves are augmented and expanded as a
linear superposition of the atom-centered solutions of the
Schroedinger equation. We name these augmented functions $\chi_\vK$,
and inside muffin-tin spheres we express them in the atom centered
coordinate system with the proper phase factor
$\chi_{\vK}(\vr)=e^{i(\vK+\vk)\vR_\mu}\bar{\chi}_\vK(\vr-\vR_\mu)$.  
For convenience of the derivation, we chose $\bar{\chi}_\vK$ to be the basis function in the muffin-tin
sphere, but without the phase factor.
The
matrix elements of the Hamiltonian are then computed by an integral of the form
\begin{widetext}
\begin{eqnarray}
\braket{\chi_{\vK'}|V|\chi_{\vK}}=\int_{int}d^3r  \widetilde{\chi}^*_{\vK'}(\vr) V(\vr) \widetilde{\chi}_\vK(\vr)
+\sum_{\mu} e^{i(\vK-\vK')\vR_\mu}\int_{MT_\mu} d^3 r\bar{\chi}^*_{\vK'}(\vr) V(\vr+\vR_\mu)\bar{\chi}_\vK(\vr)
\label{Eq:38}
\end{eqnarray}
\end{widetext}
The first term runs over interstitial space between muffin-tin (MT)
spheres, while the second term is the MT part.
We are looking for a change when we move a single atom $\mu$ at $\vR_{\mu}$ for a small amount
($\delta\vR_\mu$). The plane-wave functions $\widetilde{\chi}_\vK$ do
not change, while the augmented $\bar{\chi}_{\vK}$ in the second
integral move with the atom. In addition, because the nucleus moves,
the charge gets deformed and the potential changes for an unknown
amount $\delta V$. We will keep track of this change, but we know that
it must eventually cancel out, since we are taking derivative of a
stationary functional. This is the crucial advantage of 
a stationary functional, as otherwise one would need
to evaluate terms like 
$(\delta\Sigma/\delta G){\delta G}=(\delta^2\Phi/\delta G^2){\delta G}$, i.e., the two particle vertex
$\delta^2\Phi/\delta G^2$ would need to be computed at all frequencies, which is 
numerically extremely hard to achieve using existing impurity solvers.

Finally, we will make the usual approximation~\cite{Krakauer,W2kForce}
that the LAPW basis functions $\bar{\chi}_\vK(\vr-\vR_\mu)$ rigidly
shift with the displacement of the atom, but do not deform, in the so-called
frozen radial augmentation function approximation.

Under this assumptions, the change of a matrix elements is
\begin{widetext}
\begin{eqnarray}
\frac{\delta\braket{\chi_{\vK'}|V|\chi_\vK}}{\delta\vR_\mu}=
\braket{\chi_{\vK'}|\frac{\delta V}{\delta\vR_\mu}|\chi_{\vK}}-
\oint_{MT_\mu} d\vS\, \widetilde{\chi}^*_{\vK'}\,V\,\widetilde{\chi}_\vK
+ i(\vK-\vK')\braket{\chi_{\vK'}|V|\chi_\vK}_{MT_\mu}
+\braket{\chi_{\vK'}|\nabla V|\chi_{\vK}}_{MT_\mu}
\label{Eq:deriv}
\end{eqnarray}
\end{widetext}
The first term is due to the movement of the nucleus, and associated
change of the charge and the potential. The integral in this term is extended
over the entire space. The second term is due to the change of the
integration area for the interstitial component, and extends over the
surface of the moving MT-sphere. The third term is due to the phase
factor in Eq.~\ref{Eq:38}, and the last term arises due to the fact
that the potential in the sphere is expressed in the moving coordinate
system centered on the moving atom. 
We used here a short notation
$\braket{\chi_{\vK'}|V|\chi_{\vK}}_{MT_\mu}$ for the integral over the
MT-sphere
$\int_{MT_\mu} d^3 r \chi^*_{\vK'} V \chi_\vK$.

The matrix element for the kinetic energy operator, which takes the
form
\begin{widetext}
\begin{eqnarray}
\braket{\chi_{\vK'}|T|\chi_{\vK}}=
\int_{int}d^3r  (\nabla\widetilde{\chi}^*_{\vK'}(\vr))\cdot(\nabla\widetilde{\chi}_\vK(\vr))
+\sum_{\mu} e^{i(\vK-\vK')\vR_\mu}\int_{MT_\mu} d^3r\nabla(\bar{\chi}^*_{\vK'}(\vr))\cdot\nabla(\bar{\chi}_\vK(\vr))
\end{eqnarray}
\end{widetext}
does not have the
first and the last term of Eq.~\ref{Eq:deriv}, as the form of $\nabla\cdot\nabla$ is
originless, and hence does not change with the movement of the
nucleus, nor with the movement of the coordinate system. We thus have
\begin{eqnarray}
\frac{\delta\braket{\chi_{\vK'}|T|\chi_\vK}}{\delta\vR_\mu}=
-\oint_{MT_\mu} d\vS\, \widetilde{\chi}^*_{\vK'}\, T\,\widetilde{\chi}_\vK
\nonumber\\
+ i(\vK-\vK')\braket{\chi_{\vK'}|T|\chi_\vK}_{MT_\mu}
\label{Eq:dT}
\end{eqnarray}
Similarly, the overlap has only the following two terms
\begin{eqnarray}
\frac{\delta\braket{\chi_{\vK'}|\chi_\vK}}{\delta\vR_\mu}=
-\oint_{MT_\mu} d\vS\, \widetilde{\chi}^*_{\vK'}\widetilde{\chi}_\vK
\nonumber\\
+ i(\vK-\vK')\braket{\chi_{\vK'}|\chi_\vK}_{MT_\mu}
\label{Eq:dOlap}
\end{eqnarray}

Finally, we also need the derivative of the DMFT projector 
$\delta\left(\braket{\chi_{\vK'}|\phi_{m'}}\braket{\phi_m|\chi_{\vK}}\right)/\delta\vR_\mu$.
This can be looked at as a matrix element computed in Eq.~\ref{Eq:38},
where the potential is replaced by
$V\rightarrow \braket{\vr'|\phi_{m'}}\braket{\phi_m|\vr}=\phi_{m'}(\vr')\phi_m^*(\vr)$.
In our implementation of embedded-DMFT, the projector vanishes outside the
MT-sphere, hence the integrals over the interstitials vanishes. Inside the
MT-sphere, we rigidly shift the localized functions $\phi_m(\vr)$ and
not deform them, hence $\delta (\phi_{m'}\phi^*_m)=-\nabla
(\phi_{m'}\phi^*_m)$, so that the first and the last term in
Eq.~\ref{Eq:deriv} cancel, hence we have
\begin{eqnarray}
\frac{\delta\left(\braket{\chi_{\vK'}|\phi_{m'}}\braket{\phi_m|\chi_{\vK}}\right)}{\delta\vR_\mu}=i(\vK-\vK')\braket{\chi_{\vK'}|\phi_{m'}}\braket{\phi_m|\chi_{\vK}}
\label{Eq:derivP}
\end{eqnarray}
Note that Wannier orbitals do not rigidly shift with the atom, as
they explicitly depend on the electron charge, hence the derivative of
the projector in the Wannier basis is not so simple. Hence the Pulay
forces within the DFT+DMFT approach, implemented in Wannier basis, is
much more complicated than derived here.

Finally, let us note that the equivalent expressions 
for the derivatives Eqs.~\ref{Eq:deriv}, \ref{Eq:dT}, and ~\ref{Eq:dOlap}
were derived by Soler \& Williams~\cite{Soler_Williams}, as well as by Rici, Singh, and
Krakauer~\cite{Krakauer}. The two formalisms were shown to be
equivalent in Ref.~\onlinecite{comment_Soler}.

Next we use the Gauss theorem to simplify
\begin{eqnarray}
\braket{\chi_{\vK'}|\nabla V|\chi_\vK}_{MT} = \oint_{MT} d\vS\,  \chi^*_{\vK'}\, V\, \chi_\vK
\nonumber\\
-\int_{MT} d^3r V\, \nabla(\chi^*_{\vK'}\chi_\vK)
\end{eqnarray}
and derive a convenient expression for the change of the static part of
the Hamiltonian $H^0=T+V_{KS}$:
\begin{widetext}
\begin{eqnarray}
\frac{\delta H^0_{\vK'\vK}}{\delta\vR_\mu}=
\braket{\chi_{\vK'}|\frac{\delta V_{KS}}{\delta\vR_\mu}|\chi_{\vK}}
+ i(\vK-\vK')\braket{\chi_{\vK'}|H^0|\chi_\vK}_{MT_\mu}
-\oint_{MT_\mu} d\vS\, (\nabla\widetilde{\chi}^*_{\vK'})\cdot(\nabla\widetilde{\chi}_\vK) 
\nonumber\\
-\int_{MT} d^3r V_{KS}\, \nabla(\chi^*_{\vK'}\chi_\vK) 
+\oint_{MT} d\vS\,\left[\chi^*_{\vK'}\, V_{KS}\, \chi_\vK-\widetilde{\chi}^*_{\vK'}\,V_{KS}\,\widetilde{\chi}_\vK\right]
\label{Eq:dH0}
\end{eqnarray}
\end{widetext}
where
\begin{eqnarray}
\braket{\chi_{\vK'}|H^0|\chi_{\vK}}_{MT_\mu}=\braket{\chi_{\vK'}|T+V_{KS}|\chi_{\vK}}_{MT_\mu}
\nonumber\\
=
\int_{MT_\mu}d^3r(\nabla\chi^*_{\vK'})\cdot(\nabla\chi_{\vK}) +  \braket{\chi_{\vK'}|V_{KS}|\chi_{\vK}}_{MT_\mu}
\nonumber\\
=\braket{\chi_{\vK'}|-\nabla^2+V_{KS}|\chi_\vK}_{MT_\mu}+\oint_{MT_\mu}d\vS\, \chi^*_{\vK'} \nabla\chi_\vK
\label{Eq:TS}
\end{eqnarray}
The last term in Eq.~\ref{Eq:dH0} vanishes if the basis functions
$\chi_{\vk}$ are continuos across the MT-sphere. The continuity is
enforced in both LAPW and APW+lo method. There is however
always a very small discontinuity, which is due to the fact hat the
harmonics expansion contains finite number of spheric harmonics.  We
usually take large enough cutoff $l\approx 10$ so that this term is
around two orders of magnitude smaller than the rest of the terms, and
can therefore be safely ignored.

Next, we insert Eq.~\ref{Eq:dH0} into Eq.~\ref{Eq:Pulay}, and evaluate
term by term. The first term $\Tr(\widetilde{\rho}A^{0\dagger} \delta V_{KS} A^0)$ can be greatly simplified
\begin{eqnarray}
\Tr\left(\widetilde{\rho}A^{0\dagger} \frac{\delta V_{KS}}{\delta\vR_\mu}A^0\right)
=\sum_{ij\vK\vK'} \widetilde{\rho}_{ij}A^{0\dagger}_{j\vK'}\braket{\chi_{\vK'}|\frac{\delta  V_{KS}}{\delta\vR_\mu}|\chi_\vK}A^0_{\vK j}
\nonumber\\
=\sum_{ij} \braket{\psi_i^0|\rho|\psi_j^0}\braket{\psi_j^0|\frac{\delta  V_{KS}}{\delta\vR_\mu}|\psi_i^0}=
\Tr\left(\rho \frac{\delta  V_{KS}}{\delta\vR_\mu}\right)
\end{eqnarray}
This is because the Kohn-Sham solution $\ket{\psi_i^0}=\sum_\vK\ket{\chi_\vK}A^0_{\vK i}$
and $\widetilde{\rho}=\braket{\psi^0|\rho|\psi^0}$ is the density
matrix expressed in the Kohn-Sham basis.
Clearly this term cancels a term in Eq.~\ref{Eq:Pulay}, as expected
for stationary functional, hence the real change of the Kohn-Sham potential due
to movement of nucleus (and not due to movement of the basis attached
to the sphere) is not needed in the force calculation.

Next we simplify the forth term 
of Eq.~\ref{Eq:dH0} when inserted into Eq.~\ref{Eq:Pulay}. We have
\begin{eqnarray}
\Tr\left(\widetilde{\rho}A^{0\dagger} \int  V_{KS}\nabla(\chi^*\chi)A^0\right)
\nonumber\\
=\sum_{\vK\vK',ij}\widetilde{\rho}_{ij} A^{0\dagger}_{i\vK'} \int d^3r
  V_{KS}(\vr)\nabla(\chi^*_{\vK'}\chi_\vK) A^0_{\vK j}=
\nonumber\\
\int d^3r V_{KS}(\vr) \sum_{ij}\braket{\psi^0_i|\rho|\psi^0_j}\nabla(\psi^{0*}_{j}(\vr)\psi^0_i(\vr))
\nonumber\\
=\int d^3r V_{KS}(\vr)\nabla\rho(\vr)=\Tr(V_{KS}\nabla\rho)
\end{eqnarray}

Finally, we also simplify the last term in the Pulay forces
Eq.~\ref{Eq:Pulay}, which comes from the DMFT dynamic corrections
\begin{widetext}
\begin{eqnarray}
\vF^{dynam}\equiv -\frac{1}{\beta}\sum_{i\omega_n}\sum_{\vK\vK',m'm} \bar{G}_{\vK\vK'}(\Sigma-V_{DC})_{m'm}\frac{\delta\left(\braket{\chi_{\vK'}|\phi_{m'}}\braket{\phi_m|\chi_{\vK}}\right)}{\delta\vR_\mu}
\end{eqnarray}
\end{widetext}
Using Eq.~\ref{Eq:derivP} and the fact that the Green's function Eq.~\ref{Eq:GLAPW} can also be expressed in the
smaller Kohn-Sham basis
\begin{eqnarray}
\tilde{G}_{ij} = \left(B_{\omega_n}^R\frac{1}{i\omega_n+\mu-\varepsilon_{\vk\omega_n}} B_{\omega_n}^L \right)_{ij}=\braket{\psi^0_i|G|\psi^0_j}
\end{eqnarray}
so that $\bar{G}_{\vK'\vK} = (A^0 \tilde{G} A^{0\dagger})_{\vK'\vK}$
we arrive at
\begin{eqnarray}
\vF^{dynam}= -\frac{1}{\beta}\sum_{i\omega_n}\sum_{ij,m'm}
  \tilde{G}_{ij}(\Sigma-V_{DC})_{m'm}
\times\\
\times
\sum_{\vK\vK'}A^{0\dagger}_{j\vK'}i(\vK-\vK')\braket{\chi_{\vK'}|\phi_{m'}}\braket{\phi_m|\chi_{\vK}}A^0_{\vK i}
\nonumber
\end{eqnarray}
The projector, which expresses the DMFT Green's function in the
Kohn-Sham basis, is given by
\begin{eqnarray}
\cU_{mi}=\sum_{\vK}\braket{\phi_m|\chi_{\vK}}A^0_{\vK i}
\end{eqnarray}
from which the DMFT local Green's function is usually computed
\begin{eqnarray}
G_{loc}(i\omega_n)\equiv \cU\tilde{G}(i\omega)\cU^\dagger
\label{Eq:Glocal}
\end{eqnarray}
Note that here $G_{loc}$ is expressed in the DMFT orbital basis $mm'$.

We can compute a vector version of the DMFT projector, which is given by
\begin{eqnarray}
\vec{{\cU}}_{mi}=\sum_{\vK}\braket{\phi_m|\chi_{\vK}}\vK A^0_{\vK i}
\end{eqnarray}
to simplify the dynamic force
\begin{widetext}
\begin{eqnarray}
\vF^{dynam}= -\frac{i}{\beta}\sum_{i\omega_n}\sum_{ij,m'm}
  (\Sigma_{m'm}(i\omega_n)-V^{DC}_{m'm})(\vec{\cU}_{mi}\tilde{G}_{ij}(i\omega_n)\cU^\dagger_{jm'} -\cU_{mi}\tilde{G}_{ij}(i\omega_n)\vec{\cU}^\dagger_{jm'})
\end{eqnarray}
\end{widetext}
The first term has the form
$\Tr(\Sigma(i\omega_n)\vec{\cU}\tilde{G}(i\omega_n)\cU^\dagger)$ and
if we replace $i\omega_n\rightarrow -i\omega_n$ we get
$\Tr(\Sigma(i\omega_n)\cU\tilde{G}(i\omega_n)\vec{\cU}^\dagger)^*$,
which is complex conjugated second term. The resulting force
$\vF^{dynamic}$ is therefore a real number.

We normally compute the local Green's function by Eq.~\ref{Eq:Glocal},
but it is convenient to compute also the following vector version of
the local Green's function
\begin{eqnarray}
\vec{G}_{loc}(i\omega_n)\equiv \vec{\cU}\tilde{G}(i\omega)\cU^\dagger
\end{eqnarray}
from which the dynamic force can be computed very efficiently
\begin{eqnarray}
\vF^{dynam}= 2\Im\Tr((\Sigma(i\omega_n)-V^{DC}) \vec{G}_{loc}(i\omega_n)).
\end{eqnarray}
This calculation needs only a summation over Matsubara frequencies and
over correlated orbitals, and hence $\vF^{dynamic}$ can be computed
almost as fast as the DMFT density matrix.

Finally we insert the rest of the terms in Eqs.~\ref{Eq:dH0} and~\ref{Eq:dOlap}
into Eq.~\ref{Eq:Pulay}, to obtain the complete expression of the
Pulay forces for the valence states within LAPW basis
\begin{widetext}
\begin{eqnarray}
\vF^{Puly}_\mu =-\sum_{\vK\vK' ij}\widetilde{\rho}_{ij} A^{0\dagger}_{j\vK'}\,i(\vK-\vK')\braket{\chi_{\vK'}|H^0|\chi_\vK}_{MT_\mu}A^0_{\vK i}
\label{Eq:54}\\
+\sum_{\vK\vK'ij}\widetilde{(\rho\varepsilon)}_{ij} A^{0\dagger}_{j\vK'}i(\vK-\vK')\braket{\chi_{\vK'}|\chi_\vK}_{MT_\mu}A^0_{\vK  i}
\label{Eq:55}\\
+\sum_{\vK\vK' ij}\left[\widetilde{\rho}_{ij}  (\vk+\vK)\cdot(\vk+\vK')-\widetilde{(\rho\varepsilon)}_{ij}\right]A^{0\dagger}_{j\vK'}  A^0_{\vK i}
\oint_{MT_\mu}  d\vS\,\widetilde{\chi}^*_{\vK'}\widetilde{\chi}_\vK  
\label{Eq:56}\\
+\Tr(V_{KS}\nabla\rho)
+ 2\Im\Tr((\Sigma-V^{DC}) \vec{G}_{loc})
\label{Eq:57}
\end{eqnarray}
\end{widetext}
The first two terms contain the MT-integrals and their similar
structure but opposite sign shows how they would cancel in the absence
of the $i(\vK-\vK')$ term. The latter arises from the fact that the
basis inside MT-sphere is moved with the nucleus. Eq.~\ref{Eq:56}
contains so-called MT-surface terms which arise due to discontinuity
of the second derivative across MT-sphere~\cite{Krakauer}, and finally
the last term is due to the fact that the DMFT projector moves with
the displacement of the nucleus.

The DMFT density matrices $\widetilde{\rho}$ and
$\widetilde{(\rho\varepsilon)}$ are computed by careful summation over
the Matsubara points. Once these density matrices are computed in the
Kohn-Sham basis, we can diagonalize them
\begin{eqnarray}
&&\widetilde{\rho} \equiv \cB\; w\; \cB^\dagger
\label{Eq:271}\\
&&\widetilde{(\rho\varepsilon)} \equiv  \overline{\cB}\; (w\varepsilon) \;\overline{\cB}^\dagger 
\label{Eq:272}
\end{eqnarray}
and obtain two sets of eigenvectors $\cB$, $\overline{\cB}$ and the corresponding
eigenvalues $w_i$ and $(w\varepsilon)_i$, respectively.
Then we can insert the diagonal form for the density matrices
into Eqs.~\ref{Eq:54},~\ref{Eq:55},~\ref{Eq:56} to obtain Pulay forces
in a compact form
\begin{widetext}
\begin{eqnarray}
\vF^{Puly}_\mu =-\sum_{\vK\vK' i} w_i A^\dagger_{i\vK'}\,i(\vK-\vK')\braket{\chi_{\vK'}|H^0|\chi_\vK}_{MT_\mu}A_{\vK i}
+\sum_{\vK\vK'i} (w\varepsilon)_i \overline{A}^{\dagger}_{j\vK'}i(\vK-\vK')\braket{\chi_{\vK'}|\chi_\vK}_{MT_\mu}\overline{A}_{\vK  i}
\label{Eq:MTpart}\\
+\sum_{\vK\vK' i}\left[w_i A^\dagger_{i\vK'}  (\vk+\vK')\cdot(\vk+\vK) A_{\vK i} -
(w\varepsilon)_i \overline{A}^{\dagger}_{i\vK'}  \overline{A}_{\vK i}\right]
\oint_{MT_\mu}  d\vS\,\widetilde{\chi}^*_{\vK'}\widetilde{\chi}_\vK  
\label{Eq:inter}\\
+\Tr(V_{KS}\nabla\rho)
+ 2\Im\Tr((\Sigma-V^{DC}) \vec{G}_{loc})
\label{Eq:rest}
\end{eqnarray}
\end{widetext}
Here we used the modified eigenvectors
\begin{eqnarray}
A = A^0\cB \\
\overline{A} = A^0 \overline{\cB}
\end{eqnarray}
The resulting Eqs.~\ref{Eq:MTpart},\ref{Eq:inter},\ref{Eq:rest} have now very similar
form as the DFT Pulay forces within LAPW method~\cite{W2kForce}, except in DFT $A$ and $\overline{A}$
are both equal to the KS-eigenvectors, and $w_i$'s are fermi functions $f_i$
and $(w\varepsilon)_i$ are fermi function times KS-eigenvalues
($f_i\varepsilon_i$). The last term in Eq.~\ref{Eq:rest} bares some
resemblance to the LDA+U force~\cite{W2K_LDA+U}, but is different due to dynamic nature
of $\Sigma$ and $G_{loc}$. The algorithm to evaluate these terms is
given in appendix~\ref{AppendixA}. 

\section{Results}
\label{Chap2}

\begin{figure}
\includegraphics[width=1.0\linewidth]{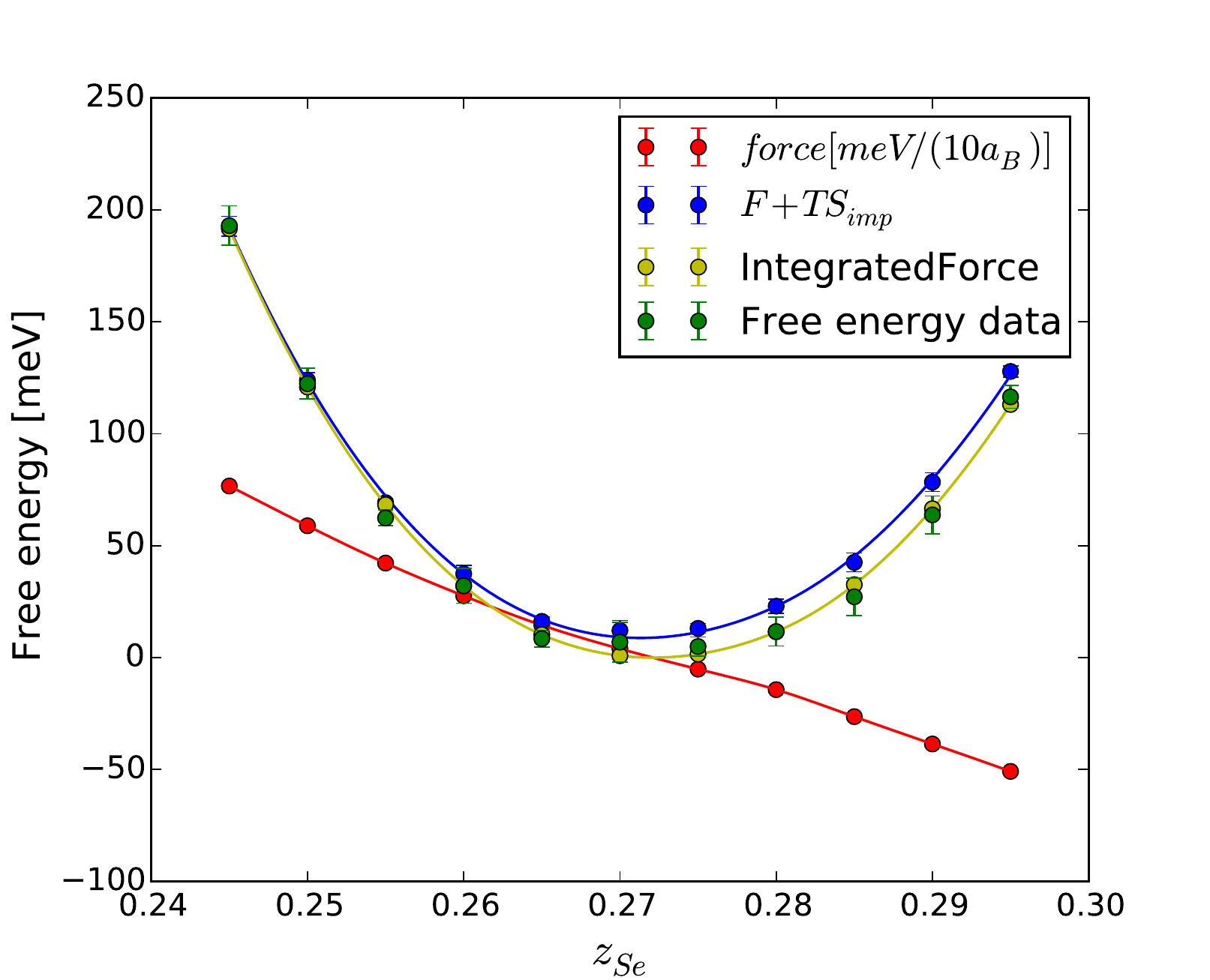}
\caption{ (Color online): Force on Se atom when displaced in
  $z$-direction, and the corresponding change of the free energy.
  The free energy is calculated from the 
  functional Eq.~\ref{DFMT:func0}, and is compared to integrated
  force.  
  We show both the free energy and $F+T S_{imp}$. The latter is directly computed in our
  method, while the former requires additional integration over the temperature.
 The quantum Monte  Carlo noise is approximately one order of magnitude smaller when
  computing energy from the force than computing it directly from the
  functional.
}
\label{fig:f1}
\end{figure}
We tested the method on several transition metal oxides, pnictides and
chalchogenides.~\footnote{These results will be published elsewhere} In this section, we show result for FeSe, one of the
most studied member of iron superconductor family, which has attracted
tremendous attention recently.
We use the implementation of DFT+EDMFT of Ref.~\onlinecite{hauleLDADMFT}, which is based
on Wien2k~\cite{wien2k}. The value of Coulomb $U$ is fixed at $5\,$eV~\cite{Kutepov}, and
we use the nominal double-counting~\cite{exactDC}.

Bulk FeSe crystalizes in tetragonal  P4/nmm structure (No. 129). It is
superconducting below 10\,K under ambient pressure~\cite{Hsu23092008}, and the
superconducting T$_c$ is increases to 37\,K under pressure~\cite{Medvedev,PhysRevB.80.064506}.  By
substitution of Se by small amounts of Te, T$_c$ can also be increased to 15\,K~\cite{Kuo,PhysRevB.79.094521},
and by intercalation with spacer layers, T$_c$ can also be boosted
to over 40\,K~\cite{FeSe_height}.

First we test the implementation of forces within DFT+EDMFTF by
computing force on Se, located at Wickoff position 2c
$(1/4,1/4,z_{Se})$ versus the Se height $z_{Se}$. As shown in
Fig.~\ref{fig:f1} the force is almost linear around the equilibrium
position, and its integral matches quite well (within the statistical
noise) to the free energy of the system.  Note that there is always
some systematic error due to frozen radial augmentation approximation,
i.e., in computing the force we do not differentiate the solutions
of the radial Schroedinger equation $u_l$.  In Fig.~\ref{fig:f1} we show both the free
energy, and the free energy without the impurity entropy. The latter
quantity is computed directly from the Green's function, while the
former needs additional integration over temperature~\cite{FreeE}. Notice that the
error-bars in computing the force are significantly smaller than the error-bars on
the free energy.

\begin{figure}
\includegraphics[width=1.0\linewidth]{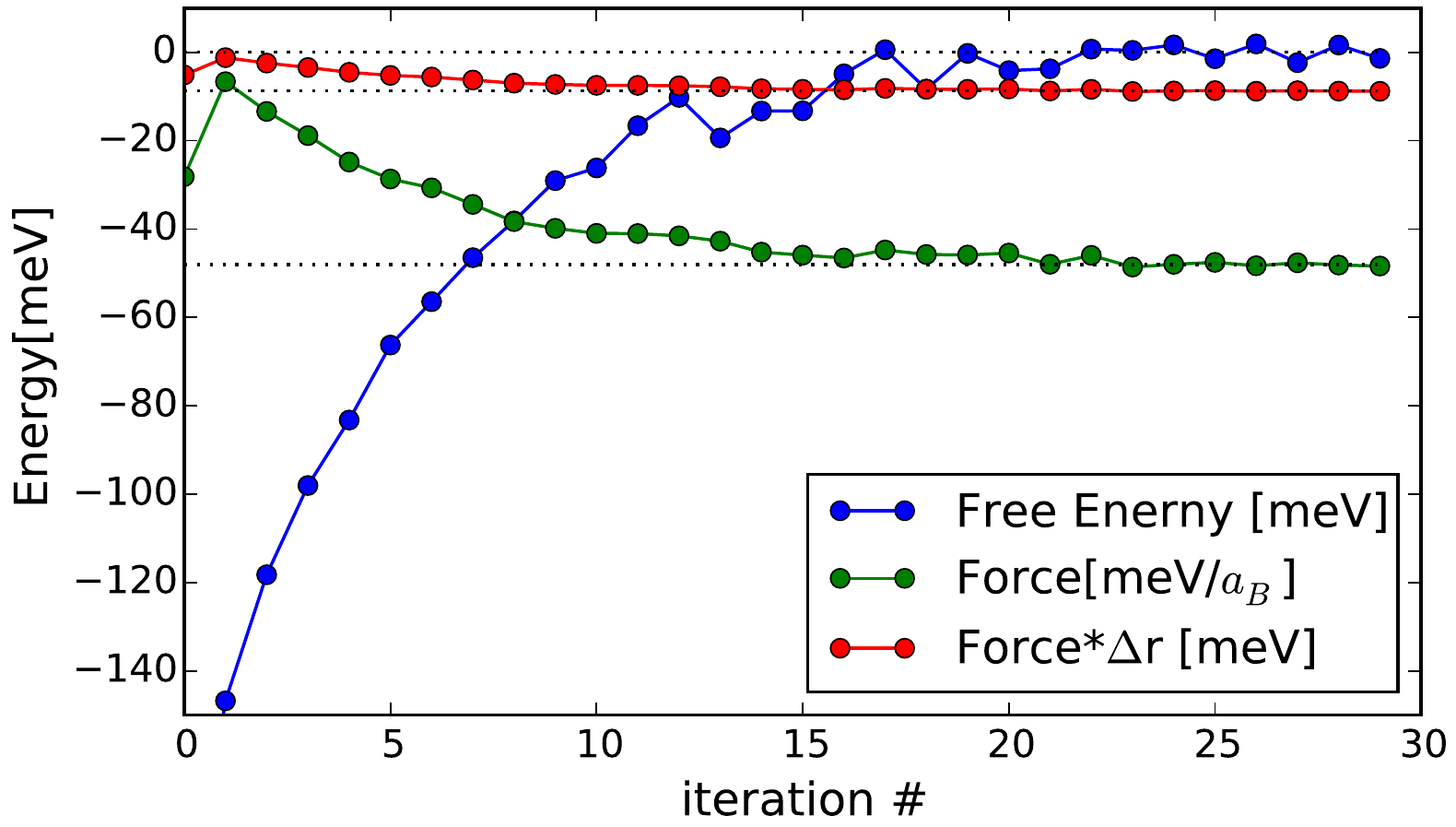}
\caption{ (Color online): The convergence of the free energy $F+T S_{imp}$
  and force with the number of DMFT iterations. 
  The last seven steps are converged, but
  display typical Monte  Carlo noise, which is more severe in free
  energy than in computing force. When the
  force is multiplied with the displacement from equilibrium $\Delta$r, to
  recover the units of energy, the noise is more than one order of
  magnitude smaller than  the corresponding noise of the free energy.
  The data corresponds to  $z_{Se}=0.25$. For clarity we subtracted a
  constant from both the energy and the force.  }
\label{fig:f2}
\end{figure}
To make this point more clear, we show in Fig.~\ref{fig:f2} the free
energy and the force from our simulation. We count as a start of the new
iteration whenever the DMFT self-energy is updated, but note that we
perform approximately 10 charge self-consistent steps for each
self-energy update, so that the charge is practically converged at
each DMFT iteration. As is clear from Fig.~\ref{fig:f2}, the Monte Carlo
noise in computing the free energy, of the order of a few meV, is present even when the free energy
is converged, and only better statistics in the QMC solver can reduce
this noise. The calculated force, measured in meV per atomic unit, has
almost factor of five smaller noise than the free energy. Finally,
when we convert the force to units of meV (by multiplying with the
distance from the equilibrium) this contribution to free energy has
almost no visible noise (approximately two orders of magnitude smaller
noise than the free energy itself). Even when we integrate the force,
to obtain the free energy, the error remains almost one order of
magnitude smaller, compared to the error in direct calculation of the
energy. We believe that this is because the $\Phi$-functional is much
more challenging to compute precisely within Monte Carlo~\cite{FreeE},
while the derivative of $\Phi$ is the self-energy, which is very
precisely sampled by the Monte Carlo method.

Many authors suggested that Se-height plays an important role in
determining superconducting T$_c$ in
Fe-superconductors~\cite{anion_height}. 
Theoretical studies of correlations in iron superconductors showed,
that the level of correlation strength is strongly coupled to the
anion-height~\cite{Yin-nm11}, as the higher anion position increases
the distance between Fe and the anion, thereby reducing the Fe-anion
hybridization. As a consequence, the strength of the local magnetic
moment is increased and correlations are increased. This is clear from
the substitution of Se by larger Te, which increases the anion heigh,
and as a consequence, the correlation strength is increased
significantly.~\cite{Yin-nm11}. Note that this effect was recently
also confirmed experimentally.~\cite{FeSeTe}

\begin{figure}
\includegraphics[width=1.0\linewidth]{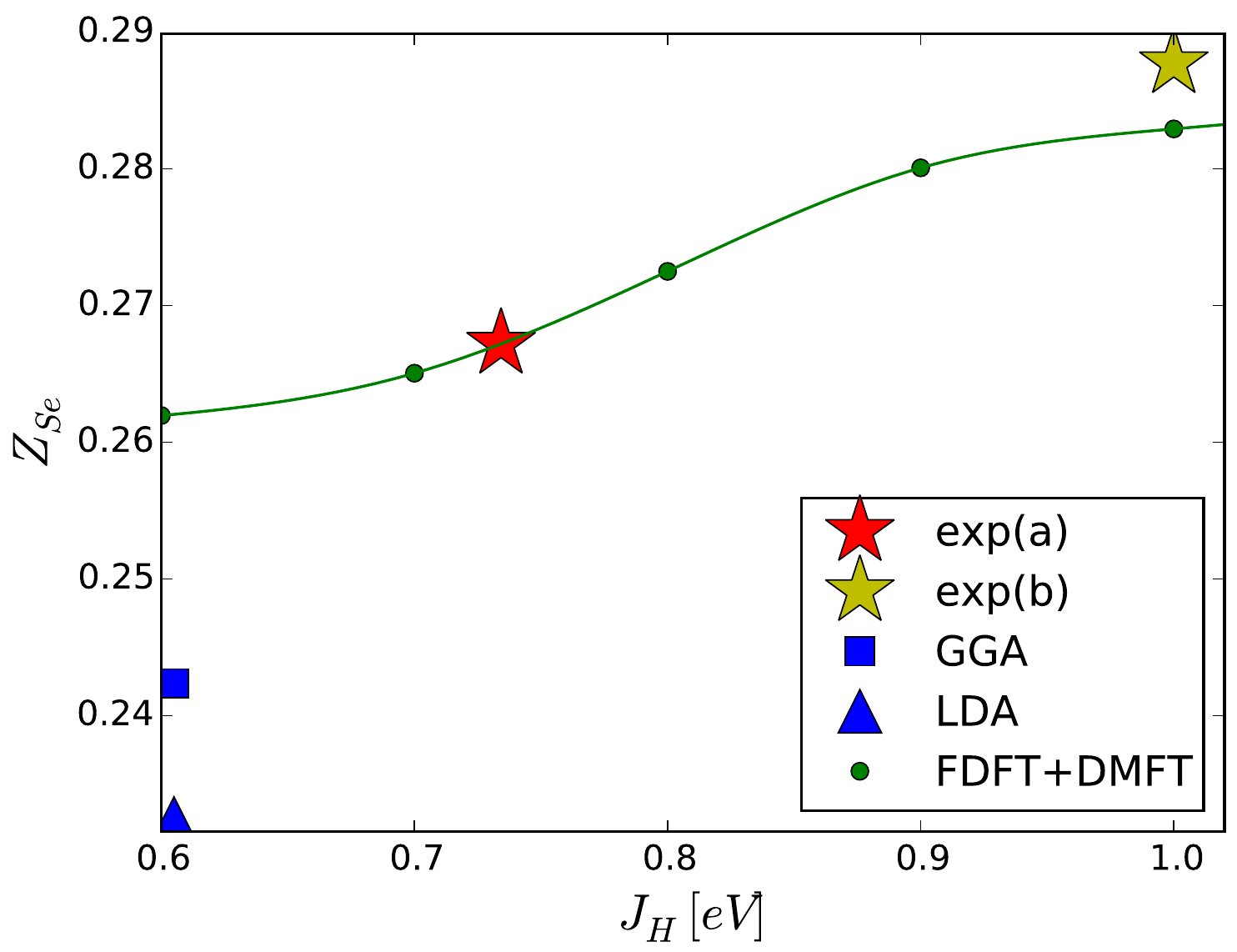}
\caption{ (Color online): 
The optimized $z$ position of Se atom for different values of Hund's
coupling $J_H$. The experimental values exp(a) and exp(b) correspond
to X-ray measurements of Ref.~\onlinecite{Cava} and
Ref.~\onlinecite{Chen}, respectively.
}
\label{fig:f3}
\end{figure}
As discussed above, previous theoretical studies and the experiments
suggest that the increased anion height leads to larger fluctuating
moment, but in the previous theoretical studies the crystal structures
of various Fe superconductors was taken from experiment, and was not
theoretically optimized.  To estimate the electron-phonon coupling in
FeSe within DFT+DMFT, the coupling between the crystal structure and
electronic structure was analyzed in Ref.~\onlinecite{CohenHaule},
using only the total energy of the system, as we did not have
implementation of forces, and structural optimization was very time
consuming.

To establish that the size of the fluctuating moment and anion height
are internally consistently predicted by the theory, one should see
that larger fluctuating moment must lead to increased anion heigh, as
otherwise cancelation effect would occur and possibly significantly
reduce or even reverse the effect, previously predicted by
theory~\cite{Yin-nm11}.

Here we calculate the optimized Se height as a function of Hund's rule
coupling $J_H$, which has a strong effect on strengthening the
fluctuating moment.~\cite{Haule_njp} It is natural to expect that an
increased fluctuating moment will reduce tendency to bind, and hence
increase anion heigh. It is however interesting to see in
Fig.~\ref{fig:f3} that this effect is strongest at exactly the
physically most relevant value of $J_H\approx 0.8\,$eV~\cite{Kutepov}. At larger
$J_H>0.9\,$eV and smaller $J_H<0.7\,$eV, the curve tends to
saturate. We thus see that FeSe is situated at exactly the critical
position, where small change of its correlation strength, or
fluctuating moment, changes its properties dramatically. It is
tempting to correlate this with experimental findings that pressure
and intercalation has a dramatic effect of its T$_c$.

We notice that both LDA and GGA significantly underestimate the
anion-height. We mark two X-ray measurements on powder samples in
Fig.~\ref{fig:f3}, which lead to somewhat different value for
$z_{Se}$. This discrepancy will likely be resolved by measurements on
a single crystal of FeSe. DMFT agrees better with Ref.~\onlinecite{Cava}, as
$J_H$ of 0.75$\,$eV is quite close to best estimates of its value in
iron superconductors~\cite{Kutepov}. The Se-heigh from
Ref.~\onlinecite{Chen} is somewhat outside the values suggested by the
present theory.  We note that Ref.~\onlinecite{Cava} considered wider range
on angles in the fit, hence it likely lead to more precise value for
$Z_{Se}$ than in Ref.~\onlinecite{Chen}.


\begin{figure}
\includegraphics[width=1.0\linewidth]{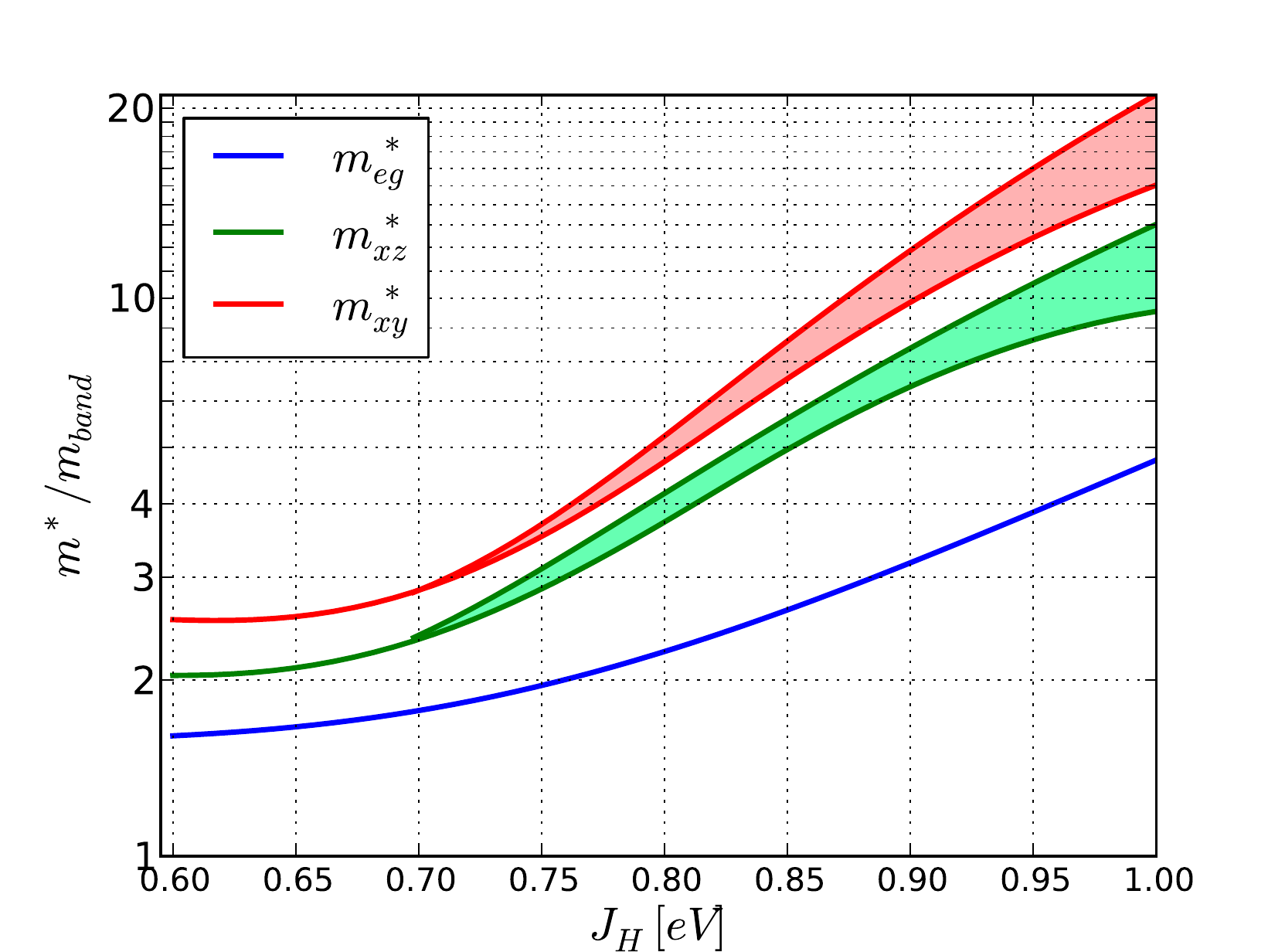}
\caption{ (Color online): 
Mass enhancement of different orbitals versus Hund's coupling $J_H$,
when $z_{Se}$ is optimized theoretically. Note logarithmic scale for mass.
}
\label{fig:f3}
\end{figure}
While the change of $z_{Se}$ from 0.265 at $J_H=0.7\,$eV to
$z_{Se}=0.28$ at $J_H=0.9\,$eV might seem small, we show below that it
has dramatic consequence for the strength of correlations on Fe atom.
Previous studies of the 5-band Hubbard model~\cite{Haule_njp} have
established that for fixed crystal structure, the increase of the
Hund's rule coupling increases effective mass and the correlation
strength. But here we show that by considering the feedback effect of
the magnetic moment on the crystal structure, this effect appears to
be even stronger. In Fig.~\ref{fig:f3} we show the strengthening of the
effective mass, as compared to LDA, for different orbitals versus
Hund's coupling.  Note that for larger $J_H$, we do not give a single
number, but rather a range of values for $m^*$. This is because our
calculation is performed at fixed temperature $T\approx 50\,$K, at
which the metallic state becomes increasingly incoherent with
increased $J_H$. In such incoherent metal, different extrapolations of
the numerical data can lead to different estimates of the mass,
hence we mark a range. The size of the spread can also be used as a
measure of incoherency, namely, as the orbital is more incoherent,
its precision for mass estimation decreases.  Experimentally, at $50\,$K the
measured band dispersion should be more consistent with the lowest
estimation of the mass, while at even lower temperature in the Fermi
liquid regime, the mass should increase and should be more consistent
with the highest estimates.

Notice that the plot is logarithmic, hence Hund's coupling increases
mass exponentially for all orbitals. Notice also that the mass
differentiation is also increased exponentially, for example at
$J_H=0.9$ the $xy$ orbital has over 50\% larger mass than $xz/yz$
orbital, while at $J_H=0.7$, the $xy$ orbital is only 20\% more
massive than $xz/yz$. Hence, Hund's coupling not just increases
correlations, but rather makes differentiation between orbitals
larger.

This is one of the central elements of the physics of Hund's
metals~\cite{Haule_njp,Zhiping_PRB}, in which spin-spin Kondo coupling
turns ferromagnetic and therefore slows down spin fluctuations,
thereby increasing the effective mass of quasiparticles, while the
charge fluctuations remain very fast, and hence charge is not blocked,
unlike in the Hubbard or t-J model. Due to coupling of the spin and
orbital through Kondo physics, the system becomes Fermi liquid at
zero temperature.~\cite{Zhiping_PRB} This physics is thus very
different from the Hubbard physics.
%


Here we used rotationally invariant Slater form of the Coulomb
interaction, where Slater integrals are related to $J_H$ by $F^2 = 8.6154 J_H$ and
$F^4 = 5.3846 J_H$. Note that the same value of $J_H$, using simpler
Kanamori parametrization of the Coulomb repulsion, leads to even larger
mass enhancements.

Note also that we do not see spin-frozen ground state, or proximity to
a quantum critical points, as found in some model studies~\cite{Gull},
whenever we use rotationally invariant form of the Hund's
coupling. When we use the density-density interaction only, which is
not rotationally invariant, we do however find spin-freezing and
incoherent metal, in which coherence is not restored with decreasing
temperature. The latter seems to be a property of certain forms of
Coulomb interactions, which do not explicitly obey rotational
invariance, and the reason behind deserves further study.

\section{Conclusions and Discussion}

In this manuscript we derived forces on atoms within ab-initio
approach termed DFT+Embedded DMFT functional. This method combines the
DFT with the DMFT such that it embeds the DMFT Feynman diagrams
directly in real space to the DFT real space functional. The resulting
functional is stationary, as we ensure that the projector 
$P=\sum_{\alpha\beta}\ket{\phi_\alpha}\bra{\phi_\alpha}\otimes\ket{\phi_\beta}\bra{\phi_\beta}$
is independent of the electronic charge density, so that $\delta
P/\delta G=0$. This property of the projector ensures that the
variation of functional $\delta\Gamma[G]$ vanishes when the 
usual Dyson Eq.~\ref{Eq:Dysn} is satisfied. Note that when Wannier
functions are used for projector, then $\delta P/\delta G$ does not vanish,
and hence the variation of the functional $\Gamma[G]$ does not lead to
a usual form of the Dyson equation Eq.~\ref{Eq:Dysn}. More complicated
Dyson equation would than need to be used.

The derivative of the stationary functional with respect to atomic
displacement was derived analytically, and we showed that the Pulay
force contains only simple terms, which appear due to our choice of
atom centered basis. We show explicitly that quantities, which are
numerically difficult to evaluate, cancel out. In particular, the two
particle vertex function, which appears due to variation of the
self-energy $\delta\Sigma/\delta G$, cancels out. Moreover, the
$\Phi[G]$ functional, which is needed for free energy evaluation, is
not needed for computing forces. The resulting forces on atoms can
thus be very efficiently computed, and we implemented them in LAPW
basis. We showed that even though quantum Monte Carlo leads to
considerable noise in evaluating the free energy (noise of the
order of a $meV$) the force contains less noise (of the order of
$0.2~meV/a.u.$), hence this precision of the force allows one to
efficiently optimize crystal structures.

We optimized the crystal structure of FeSe for different values of
Hund's coupling, and we showed that stronger fluctuating moment leads
to increase of the Se-height. The latter has dramatic impact on the
correlations in this system, as the mass increases exponentially with
the strength of the Hund's coupling. At the same time, the
orbital differentiation also increases exponentially with $J_H$. This
is the central property of the Hund's metals~\cite{Haule_njp}.

The new formula for evaluating forces on all atoms in the unit cell
within DFT+DMFT formalism thus has a great potential for both the
structural predictions, as well as prediction of phase diagrams of
correlated materials at finite temperature, which are known to have
very complex phase diagrams.

\section{Acknowledgement}
This work was supported by Simons foundation under project "Many Electron
Problem'', and by NSF-DMR 1405303.  This research used resources of
the Oak Ridge Leadership Computing Facility at the Oak Ridge National
Laboratory, which is supported by the Office of Science of the US
Department of Energy under Contract No. DE-AC05-00OR22725. We are
grateful to Gabriel Kotliar for numerous fruitful discussions, and for
carefully reading the manuscript.

\newpage
\appendix
\section{Details of the force evaluation in the LAPW basis set}
\label{AppendixA}

First we set up the notation for the LAPW basis set. The
basis functions in the interstitials are
\begin{eqnarray}
\chi_{\vk+\vK}(\vr) = \frac{1}{\sqrt{V}} e^{i(\vk+\vK)\vr}
\label{eq:bI}
\end{eqnarray}
and in the MT-spheres they take the form
\begin{widetext}
\begin{eqnarray}
&& \chi_{\vk+\vK}(\vr) = \sum_{lm,\mu}(a_{lm\mu\vK}u_l\left(|\vr-\vr_\mu|) +b_{lm\mu\vK}\dot{u}_l(|\vr-\vr_\mu|)\right) Y_{lm}(R_{\mu}(\vr-\vr_\mu))
\label{Eq:MTc}\\
&& \chi_{\nu}(\vr) = \sum_{m'\mu'}(a^{lo}_{\nu,m'\mu'} u_l(|\vr-\vr_{\mu'}|) + b^{lo}_{\nu,m'\mu'}\dot{u}_l(|\vr-\vr_{\mu'}|) + c^{lo}_{\nu,m'\mu'} u^{LO}_l(|\vr-\vr_{\mu'}|)) Y^*_{lm'}(R_{\mu'}\vr)
\label{Eq:MTLO}
\end{eqnarray}
\end{widetext}
where Eq.~\ref{Eq:MTc} stands for augmented plane wave functions,
which are matched with the plane wave Eq.~\ref{eq:bI} at the MT-sphere
boundary, and Eq.~\ref{Eq:MTLO} are additional local orbitals, which
vanish at the MT-boundary and hence do not need augmentation in the interstitials.
The index $\nu$ of the local orbitals comprises several indices
$\nu=(i_{sort},l,j_{lo},\mu,m)$, where $i_{sort}$ and $\mu$ are the
type of atom and the index of atom of a give $i_{sort}$ type,
respectively. $j_{lo}$ is the successive index of the local orbital
(as several local orbitals per atom are possible), and $l,m$ is the
index of the spherical harmonics.  Notice that in Eq.~\ref{Eq:MTLO} we sum
over all equivalent atoms $\mu'$ in the unit cell, hence a given local
orbital has a contribution in each equivalent atom and for each $m$ of
a given $l$. The precise form of the coefficients
appearing in these two equations is
\begin{eqnarray}
a_{lm\mu\vK} \equiv \bar{a}^{\vk+\vK}_{l}\;\frac{4\pi i^l S^2}{\sqrt{V}}e^{i(\vk+\vK)\vr_\mu}  Y^*_{lm}(R_\mu(\vk+\vK))
\nonumber\\
b_{lm\mu\vK} \equiv \bar{b}^{\vk+\vK}_{l}\;\frac{4\pi i^l S^2}{\sqrt{V}}e^{i(\vk+\vK)\vr_\mu}  Y^*_{lm}(R_\mu(\vk+\vK))
\label{Eq:almblm}\\
a^{lo}_{\nu,m\mu}\equiv a^{lo}_\nu\;\frac{4\pi i^l S^2}{\sqrt{V}}e^{i(\vk+\vK_\nu)\vr_\mu}  Y^*_{lm}(R_\mu(\vk+\vK_\nu))
\nonumber\\
b^{lo}_{\nu,m\mu}\equiv b^{lo}_\nu\;\frac{4\pi i^l S^2}{\sqrt{V}}e^{i(\vk+\vK_\nu)\vr_\mu}  Y^*_{lm}(R_\mu(\vk+\vK_\nu))
\nonumber\\
c^{lo}_{\nu,m\mu}\equiv c^{lo}_\nu\;\frac{4\pi i^l S^2}{\sqrt{V}}e^{i(\vk+\vK_\nu)\vr_\mu}  Y^*_{lm}(R_\mu(\vk+\vK_\nu))
\label{Eq:aloblo}
\end{eqnarray}
where $a_{lm}$ and $b_{lm}$ are determined such that the wave function
$\chi_{\vK}$ and its radial derivative are continuous across the
MT-boundary, which leads to the following set of equations
\begin{eqnarray}
\bar{a}^{\vk+\vK}_{l}=\dot{u}_l(S)\frac{d j_l(|\vk+\vK|S)}{dr} -\frac{d \dot{u}_l(S) }{dr} j_l(|\vk+\vK|S)
\nonumber\\
\bar{b}^{\vk+\vK}_{l}=\frac{d u_l(S) }{dr} j_l(|\vk+\vK|S)- u_l(S) \frac{d  j_l(|\vk+\vK|S)}{dr}
\label{eq:alm}
\end{eqnarray}
while the local orbital coefficients $a^{lo}$, $b^{lo}$, $c^{lo}$ are
determined such that the local orbital
$u^{loc}(r)=a^{lo}u_l(r)+b^{lo}\dot{u}_l(r)+c^{lo}u^{LO}(r)$ and its
radial derivative vanish at the MT sphere boundary, and the orbital is
normalized, i.e., $u^{loc}(S)=0$, $du^{loc}(S)/dr=0$,
$\braket{u^{loc}|u^{loc}}=1$.  Notice that the local orbitals
coefficients Eq.~\ref{Eq:aloblo} are given a phase factors
$e^{i(\vk+\vK_\nu)}$ in the same form as augmented waves have
(Eq.~\ref{Eq:almblm}), although local orbitals are not continued into
interstitials. The choice of momentum $\vK_\nu$ is arbitrary here, but
it is usually chosen to be a unique reciprocal vector for each local
orbital $\nu$.

\subsection{The muffin-tin term}
\label{App:a}

The potential in the MT-spheres can be divided into radial symmetric
part $V_{sym}$ and the rest $V_{nsym}$. The symmetric part of the
Hamiltonian $H^0_{sym}=T+V_{sym}$ can be compactly expressed by
\begin{widetext}
\begin{eqnarray}
(A^\dagger \braket{\chi|H^0_{sym}|\chi}A)_{MT_\mu}=
\sum_{lm}
\left(
\begin{array}{c}
\sum_{\vK'} A^\dagger_{i\vK'} a^*_{lm\mu\vK'}+\sum_{\nu}A^\dagger_{i\nu}a^{lo\;*}_{\nu,m\mu}\\
\sum_{\vK'} A^\dagger_{i\vK'} b^*_{lm\mu\vK'}+\sum_{\nu}A^\dagger_{i\nu}b^{lo\;*}_{\nu,m\mu}\\
\sum_{\nu}A^\dagger_{i\nu}c^{lo\;*}_{\nu,m\mu}
\end{array}
\right)\cH^0
\left(
\begin{array}{c}
\sum_{\vK} a_{lm\mu\vK}A_{\vK j}  +\sum_{\nu}a^{lo}_{\nu,m\mu}A_{\nu j}\\
\sum_{\vK} b_{lm\mu\vK}A_{\vK j}  +\sum_{\nu}b^{lo}_{\nu,m\mu}A_{\nu j}\\
\sum_{\nu}c^{lo}_{\nu,m\mu}A_{\nu j}
\end{array}
\right)
\end{eqnarray}
where $\cH^0=\cH^H+\cH^S$ is the sum of the volume and the surface
contribution. The volume part comes from the radial integral
$\braket{u|H^0_{sym}|u}$ and is explicitly given by
\begin{eqnarray}
\cH^{H} = \left(
\begin{array}{ccc}
E_l & \frac{1}{2} & \frac{E_l+E'_l}{2} \braket{u_l|u_l^{LO}} \\
\frac{1}{2} & E_l\braket{\dot{u}|\dot{u}} & \frac{E_l+E'_l}{2} \braket{\dot{u}|u^{LO}} +\frac{1}{2}\braket{u_l|u_l^{LO}}\\
\frac{E_l+E'_l}{2} \braket{u_l|u^{LO}_l}&\frac{E_l+E'_l}{2}\braket{\dot{u}_l|u^{LO}_l}+\frac{1}{2}\braket{u_l|u_l^{LO}}& {E'}_l\braket{u_l^{LO}|u_l^{LO}}
\end{array}
\right)
\end{eqnarray}
Here $E_l$ is the linearization energy at which the radial
Schroedinger equation is solved for $u_l(r)$, namely,
$H^0_{sym}\ket{u_l}=E_l\ket{u_l}$, and $E_l'$ is the linearization
energy of  $u_l^{LO}$, i.e.,
$H^0_{sym}\ket{u_l}=E_l'\ket{u_l^{LO}}$. The energy derivative
$\dot{u}_l$ is obtained by differentiating the above Schroedinger
equation, and takes the form $H^0_{sym}\ket{\dot{u}_l}=E_l\ket{\dot{u}_l}+\ket{u_l}$

The surface contribution comes from the fact that inside MT-sphere we
used kinetic energy operator of the form $-\nabla^2$, and in the
interstitials we used $\nabla\cdot\nabla$, which requires a surface term, as
derived in Eq.~\ref{Eq:TS}. Explicit calculation gives
\begin{eqnarray}
{\cal H}^S=S^2
\left(\begin{array}{ccc}
[u_l\frac{d u_l}{dr}]_{r=S} & \frac{1}{2}[u_l\frac{d\dot{u}_l}{dr} +\dot{u}_l\frac{d u_l}{dr}]_{r=S} & \frac{1}{2}\left[u_l\frac{du_l^{LO}}{dr}+u^{LO}_l\frac{du_l}{dr}\right]_{r=S}\\
\frac{1}{2}[u_l\frac{d\dot{u}_l}{dr} +\dot{u}_l\frac{d u_l}{dr}]_{r=S} & [\dot{u}_l \frac{d\dot{u}_l}{dr}]_{r=S} &\frac{1}{2}\left[\dot{u}_l\frac{du_l^{LO}}{dr}+u^{LO}_l\frac{d\dot{u}_l}{dr}\right]_{r=S}\\
\frac{1}{2}\left[u_l\frac{du_l^{LO}}{dr}+u^{LO}_l\frac{du_l}{dr}\right]_{r=S}&\frac{1}{2}\left[\dot{u}_l\frac{du_l^{LO}}{dr}+u^{LO}_l\frac{d\dot{u}_l}{dr}\right]_{r=S}& [u^{LO}_l\frac{d u^{LO}_l}{dr}]_{r=S}
\end{array}
\right)
\end{eqnarray}
The overlap term in the MT-sphere is computed by
\begin{eqnarray}
(A^\dagger\braket{\chi|\chi}A)_{MT_\mu}=
\sum_{lm}
\left(
\begin{array}{c}
\sum_{\vK'} A^\dagger_{i\vK'} a^*_{lm\mu\vK'}+\sum_{\nu}A^\dagger_{i\nu}a^{lo\;*}_{\nu,m\mu}\\
\sum_{\vK'} A^\dagger_{i\vK'} b^*_{lm\mu\vK'}+\sum_{\nu}A^\dagger_{i\nu}b^{lo\;*}_{\nu,m\mu}\\
\sum_{\nu}A^\dagger_{i\nu}c^{lo\;*}_{\nu,m\mu}
\end{array}
\right)\cO
\left(
\begin{array}{c}
\sum_{\vK} a_{lm\mu\vK}A_{\vK j}  +\sum_{\nu}a^{lo}_{\nu,m\mu}A_{\vK_\nu j}\\
\sum_{\vK} b_{lm\mu\vK}A_{\vK j}  +\sum_{\nu}b^{lo}_{\nu,m\mu}A_{\vK_\nu j}\\
\sum_{\vK_\nu}c^{lo}_{\nu,m\mu}A_{\nu j}
\end{array}
\right)
\end{eqnarray}
where the overlap $\braket{u|u}$ is given by
\begin{eqnarray}
\cO = \left(
\begin{array}{ccc}
1 & 0 & \braket{u_l|u_l^{LO}} \\
0 & \braket{\dot{u}|\dot{u}} & \braket{\dot{u}|u^{LO}}\\
\braket{u_l|u^{LO}_l}&\braket{\dot{u}_l|u^{LO}_l}& \braket{u_l^{LO}|u_l^{LO}}
\end{array}
\right)
\end{eqnarray}
We next carry out the expensive summation over all basis set functions
($\vK$,$\nu$) to obtain coefficients related to the band index $i$:
\begin{eqnarray}
\left(
\begin{array}{c}
a_{i,lm\mu}\\
b_{i,lm\mu}\\
c_{i,lm\mu}
\end{array}
\right)
\equiv
\left(
\begin{array}{c}
\sum_{\vK} a_{lm\mu\vK}A_{\vK j}  +\sum_{\nu}a^{lo}_{\nu,m\mu}A_{\nu j}\\
\sum_{\vK} b_{lm\mu\vK}A_{\vK j}  +\sum_{\nu}b^{lo}_{\nu,m\mu}A_{\nu j}\\
\sum_{\nu}c^{lo}_{\nu,m\mu}A_{\nu j}
\end{array}
\right)
\end{eqnarray}
and similarly we also compute a vector version of these coefficients
\begin{eqnarray}
\left(
\begin{array}{c}
\vcA_{i,lm\mu}\\
\vcB_{i,lm\mu}\\
\vcC_{i,lm\mu}
\end{array}
\right)
\equiv
\left(
\begin{array}{c}
\sum_{\vK} a_{lm\mu\vK}\vK A_{\vK j} +\sum_{\nu}a^{lo}_{\nu,m\mu}\vK_\nu A_{\nu j}\\
\sum_{\vK} b_{lm\mu\vK}\vK A_{\vK j} +\sum_{\nu}b^{lo}_{\nu,m\mu}\vK_\nu A_{\nu j}\\
\sum_{\nu}c^{lo}_{\nu,m\mu}\vK_\nu A_{\nu j}
\end{array}
\right).
\end{eqnarray}
Finally, we also compute the matrix elements of the non-spherically symmetric part of the
potential 
\begin{eqnarray}
\cV_{l m,l' m'}= \int d\Omega Y^*_{l m}(\Omega) 
\left(
\begin{array}{ccc}
\braket{u_l|V_{nsym}|u_{l'}} & \braket{u_l|V_{nsym}|\dot{u}_{l'}} & \braket{u_l|V_{nsym}|u_{l'}^{LO}} \\
\braket{\dot{u}_l|V_{nsym}|u_{l'}}  & \braket{\dot{u}_l|V_{nsym}|\dot{u}_{l'}} & \braket{\dot{u}_l|V_{nsym}|u^{LO}_{l'}}\\
\braket{u^{LO}_l|V_{nsym}|u_{l'}}&\braket{u^{LO}_l|V_{nsym}|\dot{u}_{l'}}& \braket{u_l^{LO}|V_{nsym}|u^{LO}_{l'}}
\end{array}
\right)
Y_{l' m'}(\Omega)
\end{eqnarray}

With all these coefficients $a_{i,lm\nu}$ and $\vcA_{i,lm\nu}$ in
place, we can express the MT-part of the Pulay force (Eq.~\ref{Eq:MTpart}) by
\begin{eqnarray}
\vF^{Puly-MT}_\mu=
-\sum_{\vK\vK'i} w_i A^\dagger_{i\vK'}\,i(\vK-\vK')\braket{\chi_{\vK'}|H^0|\chi_\vK}_{MT_\mu}A_{\vK i}
+\sum_{\vK\vK'i} (w\varepsilon)_i \overline{A}^\dagger_{i\vK'}\,i(\vK-\vK')\braket{\chi_{\vK'}|\chi_\vK}_{MT_\mu}\overline{A}_{\vK i}=
\nonumber\\
2\sum_{\substack{i lm\\l' m'}}
w_i \Im\left(
\left(
\begin{array}{c}
a_{i,l m\mu}^*\\
b_{i,l m\mu}^*\\
c_{i,l m\mu}^*
\end{array}
\right)(\cH^0\delta_{ll'}\delta_{mm'}+\cV_{l m l' m'})
\left(
\begin{array}{c}
\vcA_{i,l' m'\mu}\\
\vcB_{i,l' m'\mu}\\
\vcC_{i,l' m'\mu}
\end{array}
\right)
\right)
-2 \sum_{ilm} (w\varepsilon)_i \Im\left(
\left(
\begin{array}{c}
\overline{a}_{i,lm\mu}^*\\
\overline{b}_{i,lm\mu}^*\\
\overline{c}_{i,lm\mu}^*
\end{array}
\right)\cO
\left(
\begin{array}{c}
\ovcA_{i,lm\mu}\\
\ovcB_{i,lm\mu}\\
\ovcC_{i,lm\mu}
\end{array}
\right)
\right)
\end{eqnarray}

\subsection{The surface term}
\label{App:b}

The surface part of the Pulay force (Eq.~\ref{Eq:inter}) is
\begin{eqnarray}
\vF^{Pulay-SF}_\mu = 
\sum_{\vK\vG i}\left[w_i A^\dagger_{i\vK-\vG}  (\vK-\vG+\vk)\cdot(\vK+\vk) A_{\vK i} -
(w\varepsilon)_i \overline{A}^{\dagger}_{i\vK-\vG}  \overline{A}_{\vK i}\right]
\oint_{MT_\mu}  d\vS \frac{e^{i\vG\vr}}{V}
\end{eqnarray}
The convolution in basis set vectors $\vK$ needs quadratic amount of
time ($O(N^2)$). By using the fast Fourier transform (FFT)
and turning it into product in real space, it takes only $N\log(N)$
time, hence it is more efficient to use FFT on
the following quantities
\begin{eqnarray}
&&\vec{X}_i(\vr)=\sum_\vK A_{\vK,i}(\vK+\vk)e^{i\vK\vr}\\
&& Y_i(\vr)=\sum_\vK \overline{A}_{\vK,i}e^{i\vK\vr}
\end{eqnarray}
The inverse FFT is then used to obtain the surface Pulay force 
\begin{eqnarray}
\vF^{Pulay-SF}_\mu =
\int \frac{d^3r}{V} \sum_i e^{-i\vG\vr}[\vec{X}_i^*(\vr) w_i \vec{X}_i(\vr)- Y_i^*(\vr)(w\varepsilon)_i Y_i(\vr)] S^2 \int d\Omega \frac{e^{i\vG\vr}}{V}\vec{e}_\vr
 \end{eqnarray}
where the surface integral over the MT-sphere is given by
\begin{equation}
\int d\Omega e^{i\vG\vr} \vec{e}_\vr={4\pi } \frac{\vG}{|\vG|}\;j_1(|\vG|S) i e^{i\vG\vR_\mu}
\end{equation}
\end{widetext}

\subsection{The density gradient term}
\label{App:c}

Finally we give formulas to compute the gradient density term in
Eq.~\ref{Eq:rest}. The three dimensional integral can be expressed in
terms of spheric harmonics components of density $\rho_{lm}$ and
Kohn-Sham potential $V_{lm}$ as

\begin{eqnarray}
\vF^{Pulay-\nabla}_\mu&\equiv& \int d^3r V_{KS}(\vr) \nabla\rho(\vr) 
\\
&=& \sum_{\substack{l m\\ l' m'}}\int_0^\infty dr r^2  V_{l'm'}(r) \frac{d\rho_{lm}(r)}{dr} 
\braket{Y_{l'm'}|\vec{e}_r|Y_{lm}}
\nonumber\\
&+&\sum_{\substack{l m\\ l' m'}}\int_0^\infty dr r^2   \frac{V_{l'm'}(r) \rho_{lm}(r)}{r} 
\braket{Y_{l'm'}|(r\nabla)|Y_{lm}}
\nonumber
\end{eqnarray}
The following matrix elements are therefore needed 
\begin{eqnarray}
\vec{I}^{(1)}_{l'm'lm} &\equiv& \braket{Y_{l'm'}|\vec{e}_r|Y_{lm}}  \\
\vec{I}^{(2)}_{l'm'lm} &\equiv& \braket{Y_{l'm'}|(r\nabla)|Y_{lm}} 
\end{eqnarray}
and can be computed using Wigner-Eckart theorem and recursion relations for
Legendre polynomials. The result is~\cite{W2kForce}
\begin{widetext}
\begin{eqnarray}
\vec{I}^{(n)}_{l'm'lm}=c_{n,l}
\left[
-a(l,m)
\left(
\begin{array}{c}
1\\
-i\\
0
\end{array}
\right)
\delta_{m'=m+1}
+a(l,-m)
\left(
\begin{array}{c}
1\\
i\\
0
\end{array}
\right)
\delta_{m'=m-1}
+2 f(l,m)
\left(
\begin{array}{c}
0\\
0\\
1
\end{array}
\right)
\delta_{m'=m}
\right]\delta_{l'=l+1}
\\
+d_{n,l}
\left[
a(l',-m')
\left(
\begin{array}{c}
1\\
-i\\
0
\end{array}
\right)
\delta_{m'=m+1}
-a(l',m')
\left(
\begin{array}{c}
1\\
i\\
0
\end{array}
\right)
\delta_{m'=m-1}
+2 f(l',m')
\left(
\begin{array}{c}
0\\
0\\
1
\end{array}
\right)
\delta_{m'=m}
\right]\delta_{l'=l-1}
\end{eqnarray}
\end{widetext}
where 
\begin{eqnarray}
a(l,m)=\sqrt{\frac{(l+m+1)(l+m+2)}{(2l+1)(2l+3)}}\\
f(l,m)=\sqrt{\frac{(l+m+1)(l-m+1)}{(2l+1)(2l+3)}}
\end{eqnarray}
and
\begin{eqnarray}
& c_{1,l} = \frac{1}{2}      & d_{1,l}=\frac{1}{2}\\
& c_{2,l}=-\frac{l}{2}       & d_{2,l}=\frac{l+1}{2}
\end{eqnarray}
Here we use spherical harmonics definition as used is classical
mechanics. Note that in quantum mechanics literature it is customary
to add additional factor $(-1)^m$, in which case the x and
the y component of $\vec{I}^{(n)}_{l'm'lm}$ change sign.

\bibliography{force.bib}

\end{document}